Mechanically induced self-propagating reactions: Fundamentals, overview, *in situ* monitoring and perspectives

**Imelda Octa Tampubolon,[1,2] Matej Baláž[1,*]**

[1]Institute of Geotechnics, Slovak Academy of Sciences

[2]Faculty of Materials, Metallurgy and Recycling, Technical University of Košice

*corresponding author: balazm@saske.sk

## Abstract

Mechanically induced self-propagating reaction (MSR) represents a promising but a rather underexplored phenomenon within mechanochemistry, a solvent-free alternative to classical solution-based chemistry, capable of yielding functional materials within a short timeframe. On the contrary to the gradual mechanochemical reactions, MSRs are sub second combustive-like processes bringing about substantial benefits such as faster product formation, time efficiency, and lower energy consumption. Even though the mechanochemical field is in the research spotlight, MSR investigation remains on the periphery. This review aims for catalyzing a renaissance in MSR fundamentals investigation by reviewing the current state-of-art. This is done via systematically reviewing the reaction systems that were reported to react via a MSR in the past twenty years, such as metal-metalloid ones (carbides, borides, silicides, and carbonitrides), metal chalcogenides, metal oxides, and other specific types of reactions. A couple of case studies are briefly reviewed for each system, and the comprehensive table follows afterward. The review also revisits the factors impacting the MSR course from milling conditions and reagent's properties point of view, showing either a direct or inverse relationship with MSR ignition time. The recent development in the *in situ* monitoring of mechanochemical reactions allowed to obtain some mechanistic insights into the MSR process, which are summarized in the last chapter.

## 1. Introduction

The research field of mechanochemistry has been experiencing a renaissance already for more than 15 years. Since its beginnings in the ancient past, and the first research groups focusing solely on it formed in the USSR in the 20th century,[1] its potential as a synthetic methodology was discovered by the Western world only about 20 years ago.[2–4] Being an alternative to classical solution-based chemistry, it utilizes tools of high-energy ball milling to obtain desired compounds in a solvent-free, scalable, sustainable, and environmentally benign manner.[5,6] It is now recognized by IUPAC,[7] was listed among 10 innovations that will change our world,[8] follows the twelve principles of green chemistry,[9] contributes significantly to the United Nations Sustainable Development Goals[10] and now has its own journal published by Royal Society of Chemistry (RSC Mechanochemistry). The networking activities in this research field were supported also by the EU Commission in the form of a COST project in the past.[11]

Research groups developing kinetic models have significantly enhanced the mechanistic understanding of mechanochemical reactions.[12–15] However, the kinetics of mechanochemical reactions are complex and, similarly to solution-based and solid-state chemistry, can follow different kinetics.[16] A specific type of reaction is mechanically induced self-propagating reaction (MSR). Being equivalent to self-propagating synthesis (SHS) induced by heat, MSR represents a less-energy-demanding option to synthesize the desired products by inducing ignition via mechanical milling. Both SHS and MSR can be considered controlled explosive events, and if the same reagents are used, similar reactivity is observed.[17] They are combustive-like processes where, after activation, the products are rapidly formed via an exothermic reaction. There is also a possibility to ignite the SHS process during heating of the mechanically activated powders (two-stage process). This process is called mechanically activated SHS (MASHS) and was recently reviewed by Rogachev.[18,19] The first part of the review gives ideas on the influence of mechanical activation on the ignition temperature ($T_{ig}$), which is usually decreased with milling. However, sufficient energy has to be delivered, and the Tig decrease is the result of both good reagents' contact and a reduction of activation energy.[19] In the second part, he discusses the influence of mechanical activation on the propagation of the reaction front (combustion wave propagation) and the maximum combustion temperature, reporting the existence of an optimum mechanical activation time to achieve maximum values in both cases.[18]

In the case of MSRs, the reagents undergo a process of mechanical activation prior to MSR ignition. As defined by Smékal in 1942,[20] It is an increase in the reaction ability of a substance because of mechanical deformation. Thus, in addition to particles becoming smaller and having a larger specific surface area, defects are also introduced into the crystal structure. This makes the treated material more reactive and more susceptible to MSR. The principle of the MSR and main facts known from the literature about its course are summarized on the example of planetary ball milling in *Figure 1*.

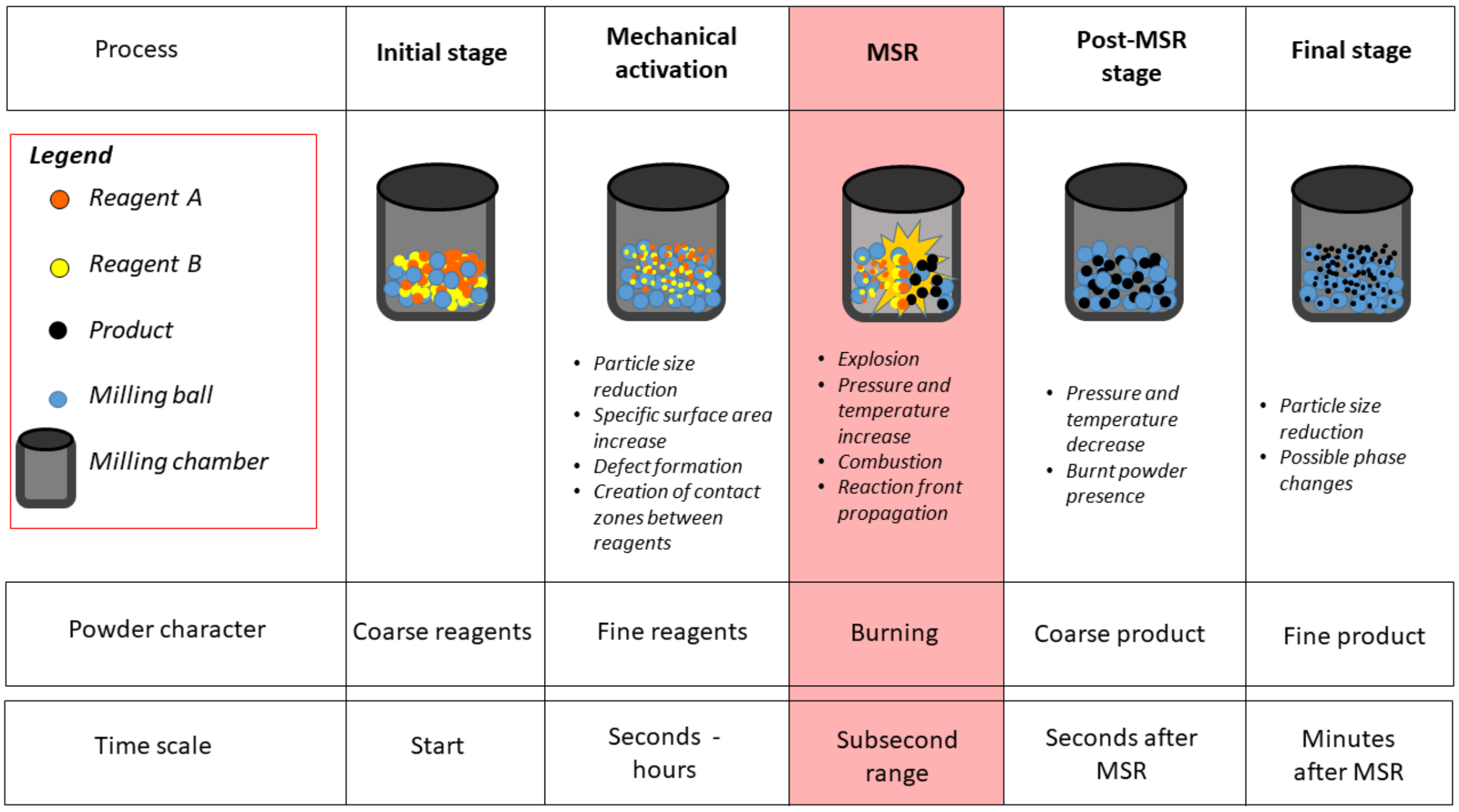

**Figure 1** Schematic illustration of MSR occurring in a planetary ball mill chamber.

To briefly address the history of MSRs, the works by Takacs represent a great source.[21,22] We have to begin with a remark made by Michael Faraday in 1820.[23] He claimed that grinding the mixture of AgCl and Zn resulted in a highly exothermic reaction yielding elemental Ag. This was later observed also for tin, copper and iron metals. The first report devoted to the systematic investigation of MSR was published by Bulgarian researchers Tschakarov, Rusanov and Bontschev in 1982 and dealt with metal chalcogenide systems.[24] The same research group kept working on the same topic, enlightening completely unknown pathways of MSRs. This pioneering work was soon followed by studies involving thermite reactions, specifically between CuO and Ca, Mg, Ti, Mn, and Fe.[25,26] In 1990, an interconnection of MSR entitled "explosive mechanochemical synthesis" by the Bulgarian researchers with the mechanical alloying field, where the same process was called "combustion synthesis by mechanical alloying," was established.[27] The scope of MSR applications was further expanded in Avvakumov et al.'s research on TiC synthesis.[28] Patankar et al.'s exploration of the metalloid system in the synthesis of $MoSi_2$,[29] and Radev et al.'s work on metal borides.[30] Since these pioneering studies, the utilization of MSR has significantly evolved, enabling the MSR observation on solid-gas and solid-liquid systems,[31] or synthesis of a broad range of compounds, e.g. nickel phosphide in 2001.[32] The only complete review devoted to MSR so far was published by Takacs in 2002.[21] In 2004, Deidda et al.[33] observed the flash of light in the chamber during MSR and were able to track the process in the subsecond range, thus bringing one of the first *in situ* insights. Research on the MSRs of metal-metalloid systems continued in the new millennium; e.g., the report by Cordoba et al. on titania carbonitrides is noteworthy.[34] Research in MSRs metal chalcogenides continued with the detailed investigation in the Zn-Se system,[35] and the discovery of the possibility of MSR also for the Cu-S system.[36] Further advancement came with the porous carbon synthesis via MSR.[37] Additionally, *in situ* investigations of MSR processes have progressed from basic observations using temperature and pressure sensors to advanced real-time studies employing spectroscopic techniques. Recently, Weidenthaler et al. utilized synchrotron techniques to study ZnS synthesis,[38] while Zheng et al. applied them to Ag nanoparticle synthesis, elucidating structural evolution and reaction mechanisms.[39] The timeline with the most crucial milestones in the MSR investigation history is provided in ***Figure 2***.

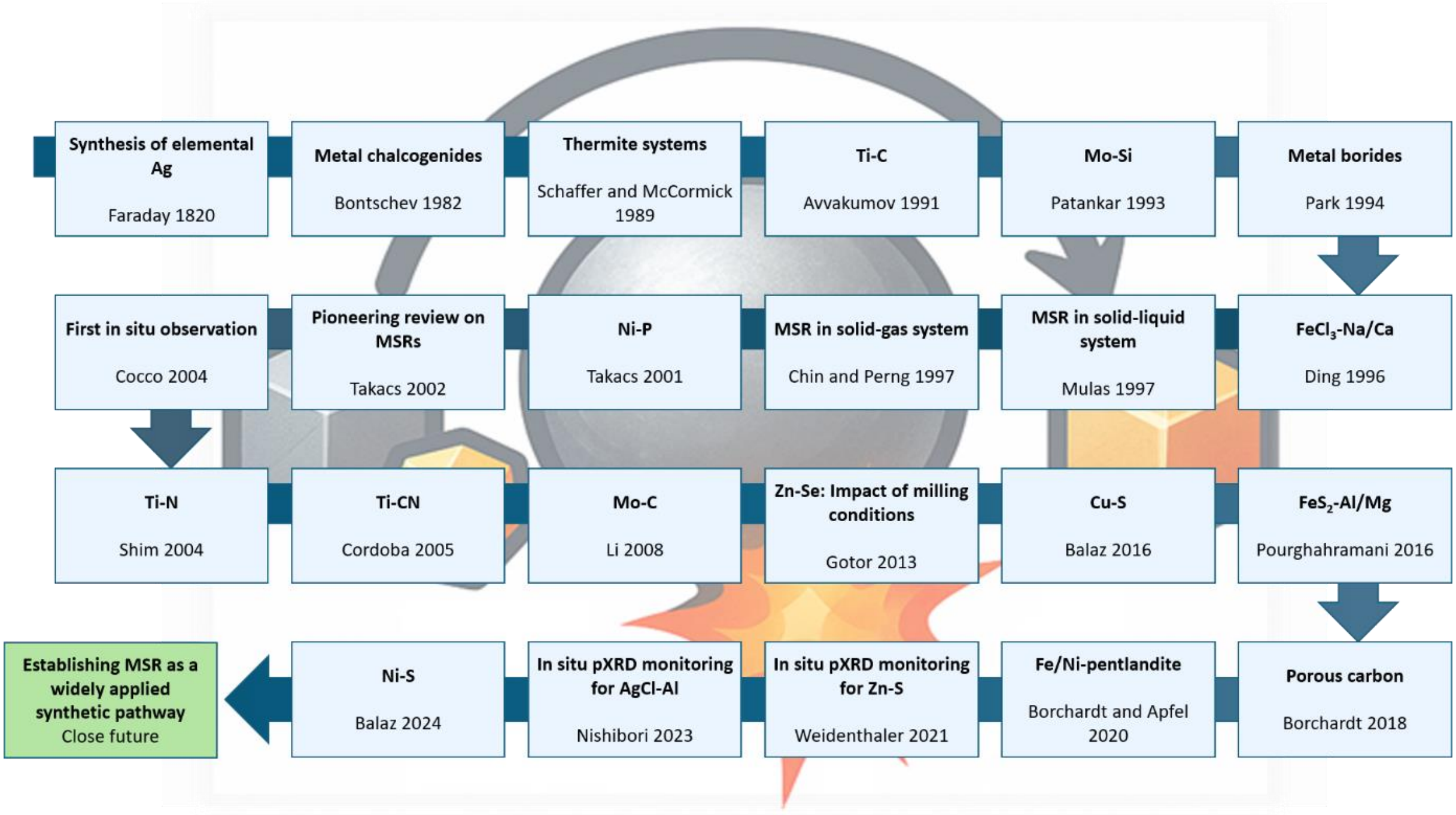

**Figure 2** Selected milestones in the history of MSR investigation. Corresponding references can be found here:[21,23,35–44,24,45,25,28,29,31–34]

Albeit mechanochemistry is in the spotlight of research nowadays, new reports on the investigation of MSRs occur only scarcely. The MSR reaction pathway is both time- and energy-saving compared to gradual synthesis, as the product can be obtained much more quickly. The MSR event is complex, and there is a large gap in understanding the fundamental processes and the influence of various input parameters. In fact, there is so little understood until now that the principles cannot be used for the rational synthesis of advanced materials.

From a thermodynamic perspective, it has long been claimed that MSRs can occur only in systems with sufficient exothermicity, where the reaction enthalpy is sufficiently negative to drive the reaction. As a rule of thumb, the $-\Delta H^{o}_{f298}/C_{p298}$, where $\Delta H^{o}_{f298}$ is the enthalpy reaction and $C_{p298}$ is the total heat capacity of the product at 298 K, commonly expressed as $-\Delta H/C$, with a minimum value of 2000 K is a criterion for the MSRs to occur.[21] When discussing potential MSRs, also adiabatic temperature, $T_{ad}$ is often discussed. Adiabatic temperature is the temperature to which the product is raised under adiabatic conditions as a consequence of the evolution of heat from the reaction.[46] In other words, it is the maximum temperature a reaction system can theoretically reach if all the heat generated by the reaction is used solely to heat the products, with no heat loss to the surroundings. A prerequisite, applicable for both SHS and MSR was that the $T_{ad}$ has to be equal to at least 1800 K for the system to become self-sustaining.[46] However, these rules have been challenged by Takacs in his review[21] and by further discoveries later on.[45,47]

In one of the early works, the impact of the hardness of the reagents was brought up.[24] However, more recent research has proven that there is more behind that, and there have to be different driving forces guiding the process.[45,47] The physicochemical and mechanical properties of the reagents, the temperature difference between them, and thermodynamic factors are being considered to impact the MSR course right now.[45,48,49] This is also because, in the past, the experimental setup did not allow for the MSR events to be reached in seconds/minutes. However, the development of milling devices led to high-energy mills with great milling efficiencies capable of igniting the MSR reactions within the second range, even for systems that were not anticipated to be possible. For MSRs to be widely applied, the scope of reactions that can be performed in this way is crucial. The periodic table in ***Figure 3*** provides an idea of which elements are being studied in relation to MSR.

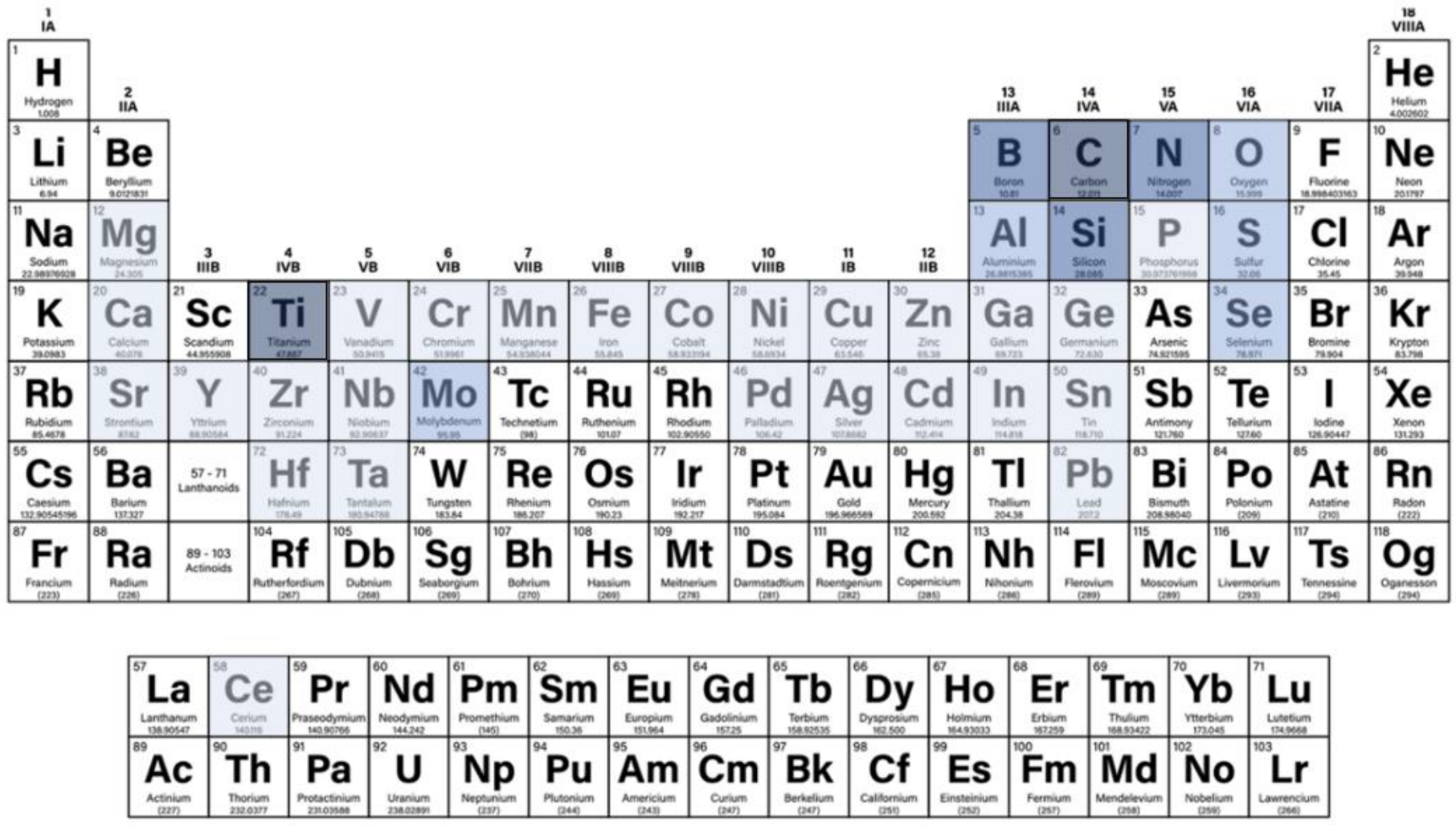

Figure 3 Periodic table with highlighted elements studied in MSR research. The darker the highlighting, the larger the number of papers devoted to the MSRs concerning the given elements and compounds containing them. The studies published in between 2002 and 2024 are considered.

The periodic table gives an impression of quite a wide scope of reactions susceptible to MSR. The most studied compound is definitely titanium carbide. In general, the most research in the field of MSRs was devoted to preparation of refractory compounds, such as borides, carbides, nitrides, etc.[50] However, this phenomenon was also reported, among others, for the synthesis of metallic nanoparticles,[39,51] phosphides,[52] organics,[53] or metal chalcogenides (MCs).[24] The major part of this review is devoted to providing the reader with an overview of various systems for which MSRs were reported in the last twenty years.

So far, there is a lack of information on how to precisely control the course of the MSR process on a general level. There are many reports reporting on the influence of milling conditions (such as the mass ratio between the milled powder and the milling balls, rotation speed of the mill, or milling time) on the ignition time of MSRs (e.g. [35,54]) and also the effect of inert additives and change in stoichiometry is often considered . Nevertheless, the observed results are usually not related to the properties of the reagents and are not generalized for a wider understanding of the MSRs fundamentals.

Research in MSR stays on the periphery of mechanochemistry despite the fact that it offers a time- and energy-saving possibility. Prof. Takacs published the only review devoted solely to the topic of MSRs in 2002 and 2014.[21,22] It is an extensive review covering various types of reactions susceptible to MSR; models used to describe MSR processes, and the influence of various factors on the MSR course.

The present review aims to provide an update to it since a considerable amount of time has already passed, and there have been some advancements in understanding the MSR process. Since then, the MSR has also been reported to occur in systems where it was never hypothesized to be possible due to not respecting a rule of thumb regarding the high enough exothermicity of the given reaction system. Moreover, the availability of *in situ* monitoring of temperature and pressure during milling is now commercially available. Also the development of time-resolved *in situ* X-ray diffraction[55] and Raman spectroscopy monitoring [56] techniques brought the research of MSR to new heights. It is the objective of this review to provide readers with a general picture of the state-of-the-art in the MSR research in terms of reactions scope, influence of milling parameters, and tools available for their monitoring and to emphasize the main challenges still waiting to be tackled in the future.

## 2. Reaction systems for which MSR was reported since 2002

In this chapter, we have systematically divided the reports into sub-chapters based on the synthesized compounds, namely the preparation of metal-metalloids, metal chalcogenides, metal oxides, and the rest of the studies. Each section is composed of a brief description of one or two case studies that we found the most interesting in terms of understanding the MSR process or are related to the system mostly studied for the given system. This is followed by a comprehensive table mentioning the works that have appeared since the publication of the pioneering review.[21]

### 2.1. Metal metalloids

Carbides, borides, and silicides, categorized as metalloid groups, are commonly referred to as hard metals. The term hard metal was first introduced by Schwarzkopf and Kieffer in 1953,[57] specifically with reference to ceramics composed of carbides, nitrides, borides, and silicides of metals in groups IV to VI of the periodic table. These metalloid compounds exhibit remarkable physicochemical properties, including high melting points, electrical conductivity, resistance to wear, hardness, and toughness. Traditionally, these compounds are synthesized through the combustion process,[58] often forming solid solutions. The pioneering mechanosynthesis of metal carbides was introduced by Matteazi and Le Caer, who successfully prepared carbides of Ti, V, Cr, Mn, Fe, Co, Ni, Zr, Nb, Mo, Ta, W, Re, Al, and Si by milling these metals with carbon at room temperature.[59] While their work was not explicitly aimed at systematically investigating the MSR, it was shown that these compounds can be prepared in this way.

### 2.1.1. Carbides

Carbides are compounds composed of metallic or semimetallic elements combined with carbon. They exhibit exceptional thermal stability, catalytic performance, and mechanical hardness, making them highly valuable in industrial applications.[60,61]. Commonly synthesized carbides via MSR method include titanium carbide,[62–64] tungsten carbide,[65,66] and vanadium carbides.[67,68] Additionally, aluminum is sometimes incorporated to form complex carbides such as TiAlC[69] and $Cr_2AlC$.[70] Also, MAX phases composed of a metal (M), an A-group element (A, most often from the group IIIA or IVA), and an element such as C, N, B, or P (X), can be prepared via MSR.[71] In the pioneering review,[21] eleven reports devoted to the MSRs of carbides are mentioned.

As already outlined in the introduction, the most studied material with regard to MSR is titanium carbide, and some insights into the MSR mechanism can be traced in a few studies devoted to this system.

Yang et al. conducted a comprehensive study on the synthesis of titanium carbide (TiC) using the MSR in a horizontal rotary ball mill, with a high milling speed of 900/1000 rpm.[62] The synthesis proceeds according to the reaction:

$$Ti + C \rightarrow TiC \quad (1)$$

The research demonstrated the scalability of TiC production by mixing Ti and C in a total mass of 100 g, achieving the successful product formation within 30 minutes of milling. Carbon gradually diffused into Ti, with the evidence that the intensity of (002) plane of graphite disappeared and the that of the (002) plane of Ti increased after 15 minutes of milling, as shown in the corresponding XRD pattern in ***Figure 4***a. The MSR was triggered in between 15 and 30 minutes, indicated by the disappearance of the diffraction of the (002) Ti plane and the emergence of sharp peaks corresponding to the TiC phase detected in the XRD pattern after 30 minutes of milling. The most significant diffraction peaks of TiC phase correspond to (111), (200), and (220) crystallographic planes. Prolonged milling led to a slight broadening and shift of the TiC peaks, signifying a reduction in crystallite size, and increase of the lattice parameters value, respectively, corroborating the stabilization of nanocrystalline TiC. The milling process influenced the particle size distribution (***Figure 4***b). Initially, the $d_{50}$ value (median particle size) of graphite and Ti powder were 8.91 µm and 20.2 µm, respectively. The $d_{50}$ value (accounting for both elements now) increased to 27.61 µm because of the deformation of Ti into a flake or platelet shape due the impact of the milling balls [72]. Upon the MSR completion before 30 minutes, the particle size rapidly decreased to 5.61 µm, indicating the formation of fine TiC particles. Further milling reduced the particle size to 0.98 µm detected after 60 min, with subsequent prolonged milling resulting in some aggregation, increasing the diameter slightly to 2.75 µm after 210 min. The TEM/HRTEM analysis in ***Figure 4***c confirmed the successful reaction of graphene with Ti and the formation of small TiC nanoparticles.

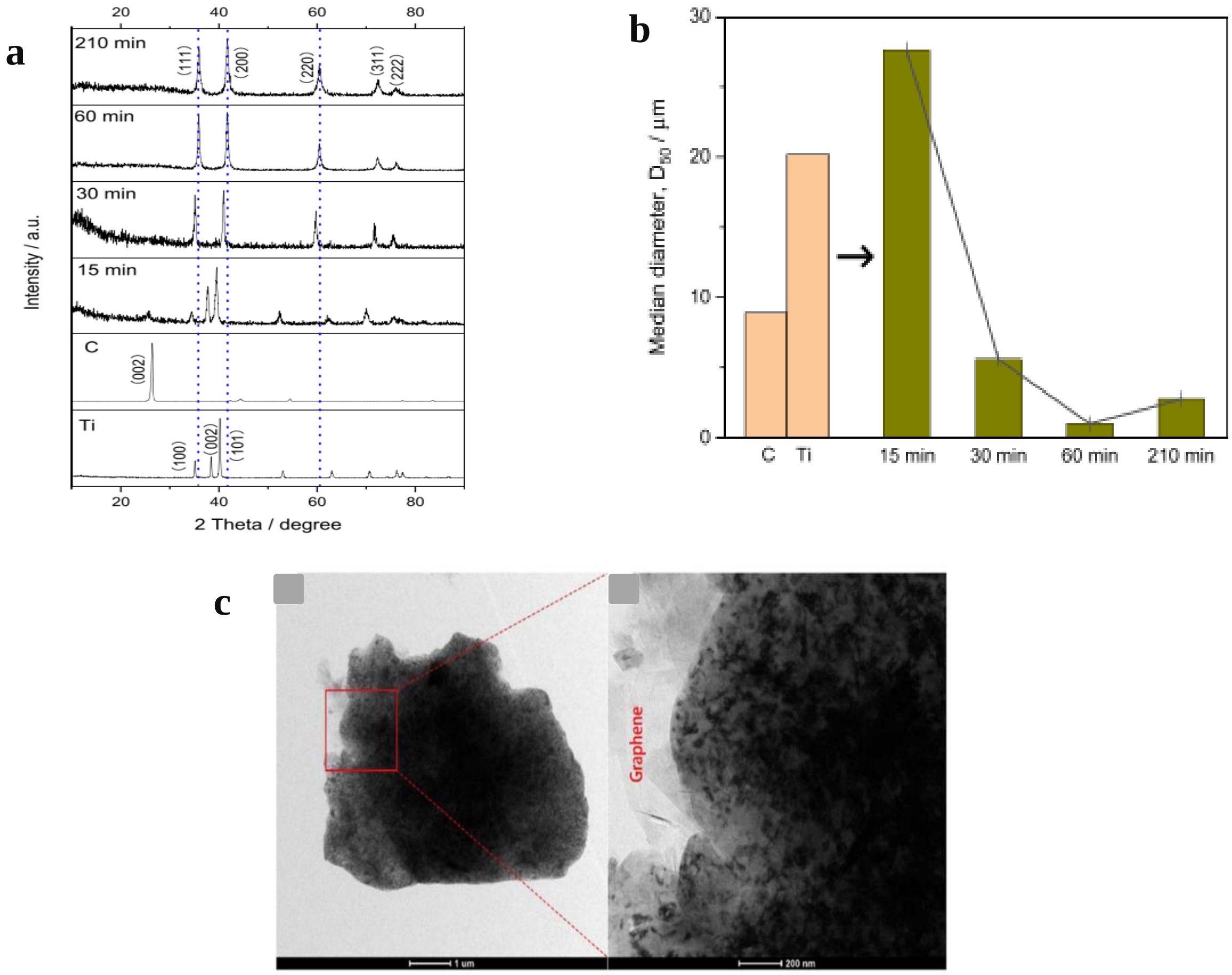

Figure 4 (a) XRD patterns of reagents and $Ti_{50}C_{50}$ mixture milled for different time, (b) d50 (median) particle size values of reagents and $Ti_{50}C_{50}$ mixture milled for different time, and (c) TEM (left) and HRTEM (right) images of the products of $Ti_{50}C_{50}$ mixture after 30 min of milling. Reprinted with permission from [62]. 2020 licensed under CC-BY.

Oghenevweta et al. studied the microstructural changes of the Ti and C reagents prior to MSR and that of the TiC product after MSR in [73]. The reaction was conducted in a magnetically controlled ball mill under a helium atmosphere. The temperature was monitored *in situ* during milling (***Figure 5***). The MSR occurred after 36 hours of milling. However, the subsequent cooling stage took a considerably long time of about 1.5 hours after the MSR. The insight into microstructural changes was provided by SEM and TEM (***Figure 6***a and b). After 25 hours of milling, metallic Ti particles were deformed, gradually evolving into a layered or pancake-like structures. As milling continued for 36 hours, graphite flakes broke into particles of various sizes, and both eventually refined into a heterogeneous mixture of Ti and graphite at the micro- and nanoscale. HRTEM images in ***Figure 6***c revealed regions composed of bent and curved graphitic flakes and flower-like particles already after 5 hours of milling. After 25 hours (***Figure 6***d), the graphite particles developed into highly twisted or nearly spherical shapes. These twisted graphitic layers appeared as loops penetrating the titanium crystal, suggesting that the graphite flakes may have clumped around the Ti particles.

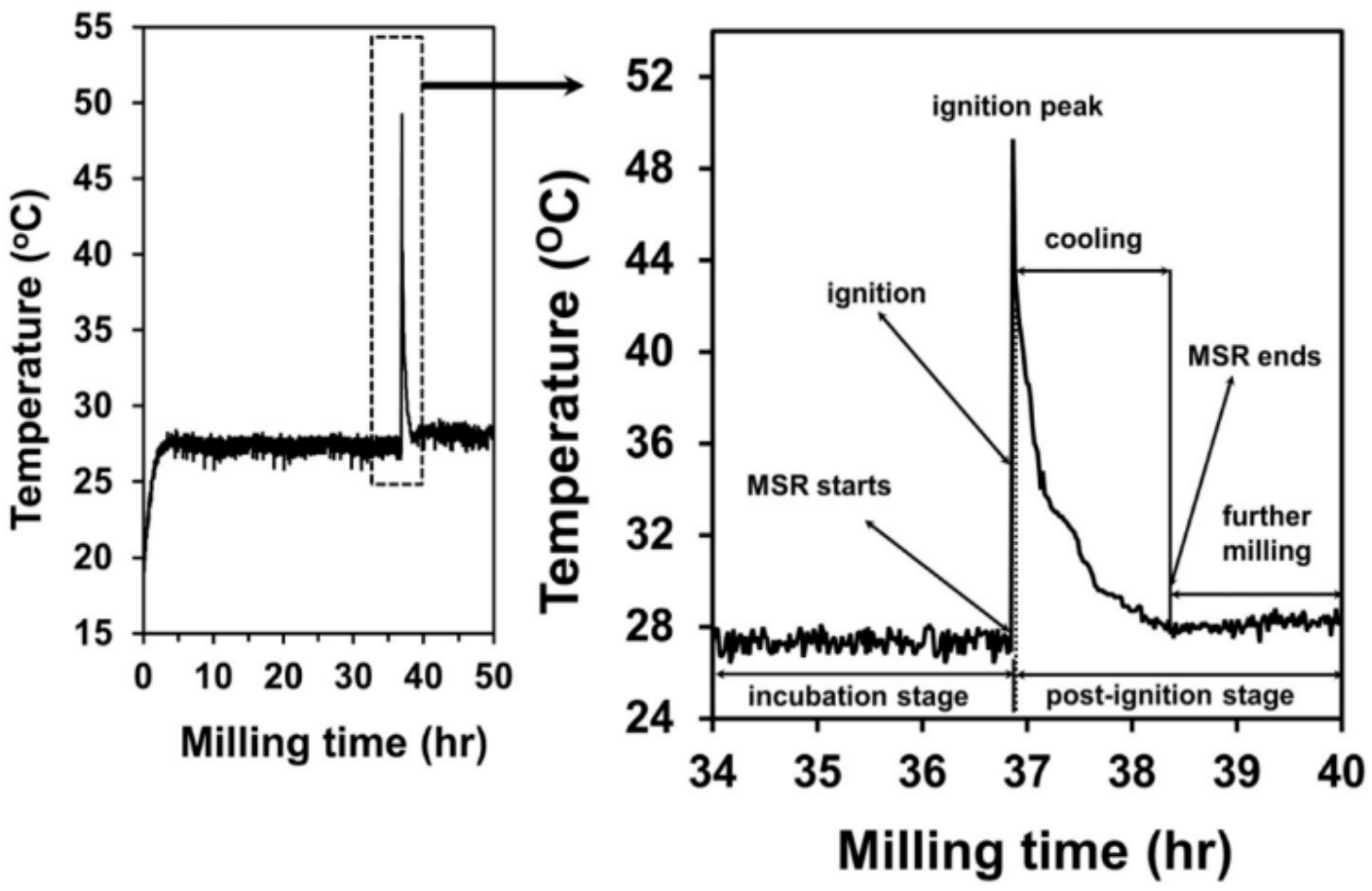

Figure 5 Plots of external temperature of the vial vs milling time during magnetically controlled ball milling of $Ti_{50}C_{50}$ mixture. Reprinted from [73] Copyright (2018), with permission from Elsevier.

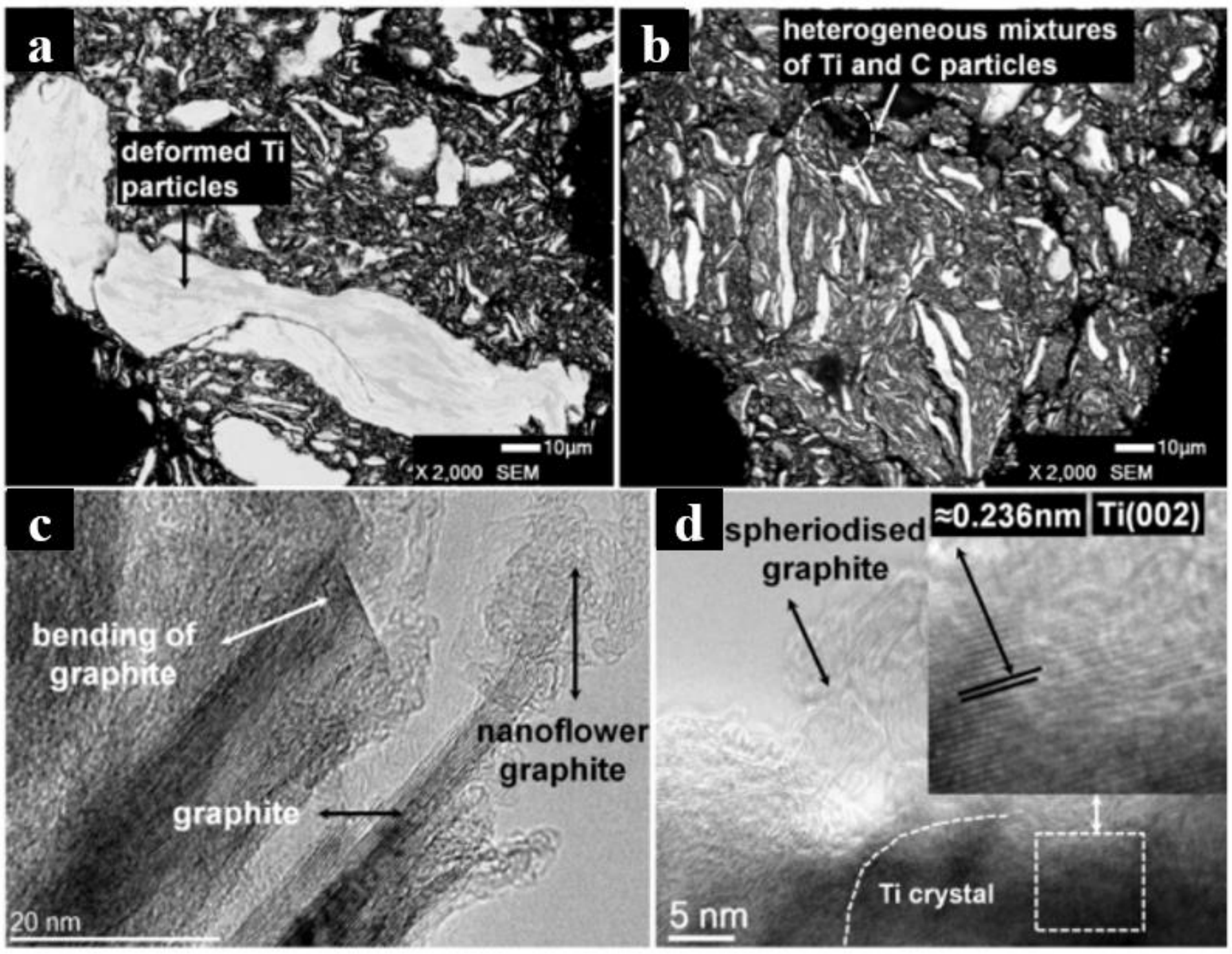

Figure 6 Microstructural analysis of the $Ti_{50}C_{50}$ mixture before MSR ignition: (a) SEM image after milling for 25 h, (b) SEM image after milling for 36 h, (c) HRTEM image after milling for 5 h, (d) HRTEM image after milling for 25 h. Reprinted from [73] Copyright (2018), with permission from Elsevier.

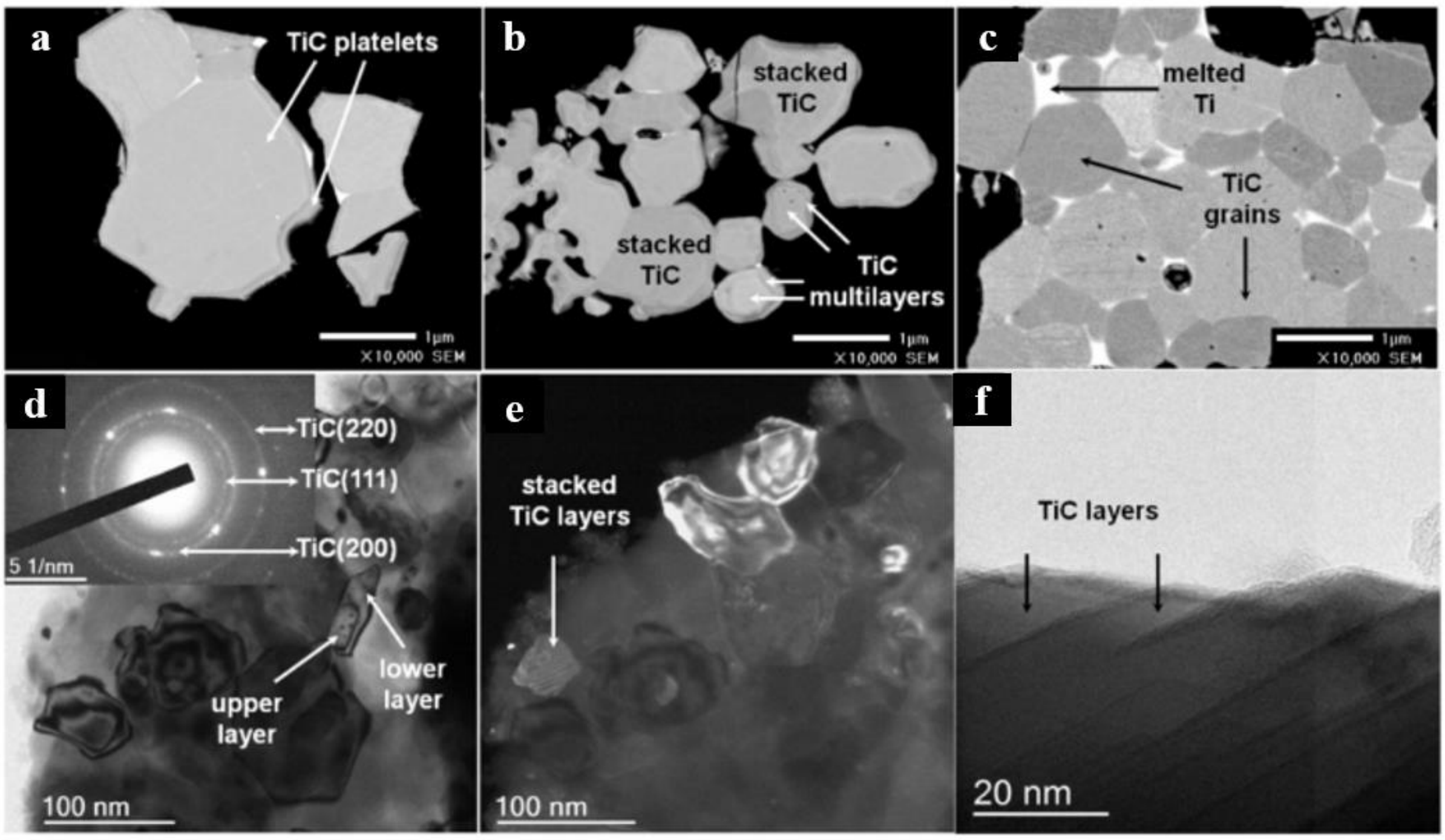

**Figure 7** Microstructural analysis of the $Ti_{50}C_{50}$ mixture after 37 h of milling, after MSR ignition (a) SEM images of TiC, (b) TiC stacks or layers, (c) TiC grains bonded by melted Ti particles, (d) bright-field TEM image of TiC grains, (e) dark-field TEM image, (f) HRTEM image displaying TiC multilayers. Reprinted from [73] Copyright (2018), with permission from Elsevier.

After the MSR, the resulting products include large TiC platelets with multiple layers that are sometimes stacked (***Figure 7***a and b). Also, coarse TiC particles bonded together by locally melted metallic titanium particles (white regions) were identified (***Figure 7***c). The local melting of Ti particles is likely due to the significant heat released during the exothermic reaction at MSR ignition, which contributes to the liquid-phase sintering of TiC, promoting rapid grain growth. The bright-field TEM images (***Figure 7***d) show the upper and lower TiC layer of the stacked structure, and the dark-field TEM image (***Figure 7***e) reveals stacks of TiC and ***Figure 7***f is the TEM image displaying the well-defined TiC multilayers.

These two case studies represent only a very minor part of the research done on MSRs of carbides. There are many other studies devoted to this topic. They are summarized in Table 1 below in terms of the reagents, target compounds, ignition time and mill type. It can be seen that the majority of MSRs were observed in a planetary ball mill, providing the largest amount of energy. Specifically for the formation of TiC via MSR, the formation of an off-stoichiometric compound $TiC_x$ is usually observed, resulting in a high proportion of carbon vacancies, as reported in [62–64,74,75].

**Table 1.** Overview of the studies devoted to MSR investigation of carbides since 2002

| Year | Reagents | Target compounds | Ignition time | Mill type | Ref. |
|---|---|---|---|---|---|
| 2024 | Ti, Al, Graphite (C) | $Ti_3AlC_2$ | 7.5 to 8.33 h (depends on Al content) | Planetary mill | [76] |
| 2022 | $NiO_2$, $Cr_2O_3$, C, Mg | Ni-$Cr_3C_2$ | 1.5 h (excess Mg) | Planetary mill | [77] |
| 2021 | Ti, Cr, C, Al, Carbon black, | $Ti_2AlC/Cr_2AlC$ | 5 min | Planetary mill | 70 |
| 2020 | Ti, C | TiC | 15 min | Horizontal rotary ball mill | 62 |
| 2019 | $CaWO_4$ + Mg+ C | WC, $W_2C$ | 38- 46 min (depends on Mg content) | Planetary mill | 65 |
| 2018 | NiO, $V_2O_5$, Mg and C | Ni-VC | 30 min and 40 min | Planetary mill | 67 |
| 2018 | Mg, $V_2O_5$, $MoO_3$, C | Mo-$V_2C$ | 12 min | Planetary mill | 68 |
| 2018 | Ti and C | TiC | 37 h | Magnetically controlled ball mill | 73 |
| 2017 | $B_2O_3$-$Cr_2O_3$+Mg+C | $CrB_2$-$Cr_2C_3$ | 4 h | Planetary mill | 78 |
| 2016 | Ti, Carbon black | Ti-C | 12 min | Planetary mill | 79 |
| 2016 | Mg, $V_2O_5$, $Co_3O_4$, C | VC-Co | 10 min | Planetary mill | 80 |
| 2016 | $TiO_2$, Al, C | TiC | 3 h | Planetary mill | 81 |
| 2016 | Ti, Graphite | Ti-C | 37 min | Magnetically controlled ball mill | 82 |
| 2016 | Ti, Al, Graphite (C) | $Ti_3AlC_2$ | 8.33 h | Planetary mill | 83 |
| 2015 | Ti, C | $Ti_{55}C_{45}$ | 12 h | High-energy ball mill | 84 |
| 2015 | $V_2O_5$, Mg, C | VC and $V_2C$ | 50 min | not mentioned | 85 |
| 2015 | Mg, $ZrO_2$, C | ZrC | 12 h | not mentioned | 86 |
| 2014 | Ti,C | $Ti_{55}C_{45}$ | 172 min | Planetary mill | 87 |
| 2014 | Ti,C | TiC | not mentioned | Mixer mill | 75 |
| 2014 | $MoO_3$, Mg, C, | Mo2C | 12 h | Planetary mill | 88 |
| 2013 | Al, $B_2O_3$, C | $B_4C$-$Al_2O_3$ | 4 h | Planetary mill | 89 |
| 2012 | Ti, Si, C | $Ti_3SiC_2$ | 3 h | Planetary mill | 90 |
| 2010 | $WO_3$, C, Mg | WC/MgO | 4.7 h | Planetary mill | 91 |
| 2010 | Ti, Al, C | $Ti_2AlC$ | 24 h | Planetary mill | 85 |
| 2009 | $WO_3$, C, Al | WC/$Al_2O_3$ | 7 min | Planetary mill | 92 |
| 2009 | Ti, Al, C | $Ti_3AlC_2$ | 3 h | Planetary mill | 69 |
| 2008 | $MoO_3$, Al, C | β-$Mo_2C$, η-$Mo_3C_2$ | Ranging from 6.5 h to 29.5 h (depends on C content) | Planetary mill | 93 |
| 2007 | Ti, C | TiC | 41 h and 71 h (depends on the Ti and C content) | Planetary mill | 92 |
| 2006 | Ti, Al, C | $Ti_3AlC_2$ | 4 h | Planetary mill | 63 |
| 2005 | Ti, Si, C | $Ti_3SiC_2$ | 1.5 h | Planetary mill | 94 |
| 2005 | Ti, C | TiC | 40 h | Magnetically controlled ball mill | 74 |
| 2005 | Ti, C | TiC | 40 h and 70 h (depends on the starting composition) | Magnetically controlled ball mill | 64 |
| 2005 | Si, Ti, C, | TiC, SiC | 72 h | Planetary mill | 95 |
| 2005 | Ti, Al, Si, C, | Ti, TiC, $Ti_5Si_3$ | 5 h | Planetary mill | [96] |

### 2.1.2. Borides

The second mostly studied metal-metalloids group in terms of MSR investigation is borides, known as refractory hard metals, and useful for the development of advanced materials for high-temperature technologies, such as in defense, nuclear, and high-temperature industries,[97] and for the catalysts for oxygen evolution reaction.[98] The synthesis of these materials is crucial for their performance in these demanding environments. Traditionally, SHS has been the standard method for producing metal borides. This method was first successfully used by Merzhanov et al.[99] to synthesize borides like $TiB_2$ and $ZrB_2$. However, the SHS process often yields a porous and hard solid mass, requiring additional milling to produce a fine powder suitable for further applications.[21]

In the pioneering review by Takacs,[21] there are six studies devoted to borides synthesized via MSR pathway. Similarly to carbides, titanium-based compounds are among the most studied compounds also in the case of borides and thus, we have selected two case studies devoted to this system.

Aviles et al.[100] investigated the synthesis of $TiB_2$–$ZrB_2$ and $TiB_2$–$HfB_2$ solid solutions under inert conditions by milling the mixtures of Ti/Zr/B and Ti/Hf/B in a planetary mill. The ignition time was contingent on the initial composition of the mixtures, and the stoichiometry of the solid solutions was adjusted by altering the atomic ratios of the components, allowing for the complete compositional range of the $TiB_2$–$ZrB_2$ systems to be achieved via MSR. Notably, $TiB_2$–$HfB_2$ solid solutions demonstrated excellent chemical uniformity, except for Ti-rich mixtures. Both chemical composition and lattice parameters exhibited a linear relationship across the systems.

Also the pure titanium-boron mixture was investigated in terms of producing titanium boride via MSR in.[101] The process was studied for two stoichiometries, to yield TiB and $TiB_2$ compounds. *In situ* monitoring of the temperature during milling allowed it to precisely detect the moment of MSR ignition. The ignition time was observed after 59 h of milling of $Ti_{50}B_{50}$ (TiB) mixture and after 18 h of milling of boron-rich $Ti_{34}B_{66}$ ($TiB_2$) mixture (***Figure 8***a and b).

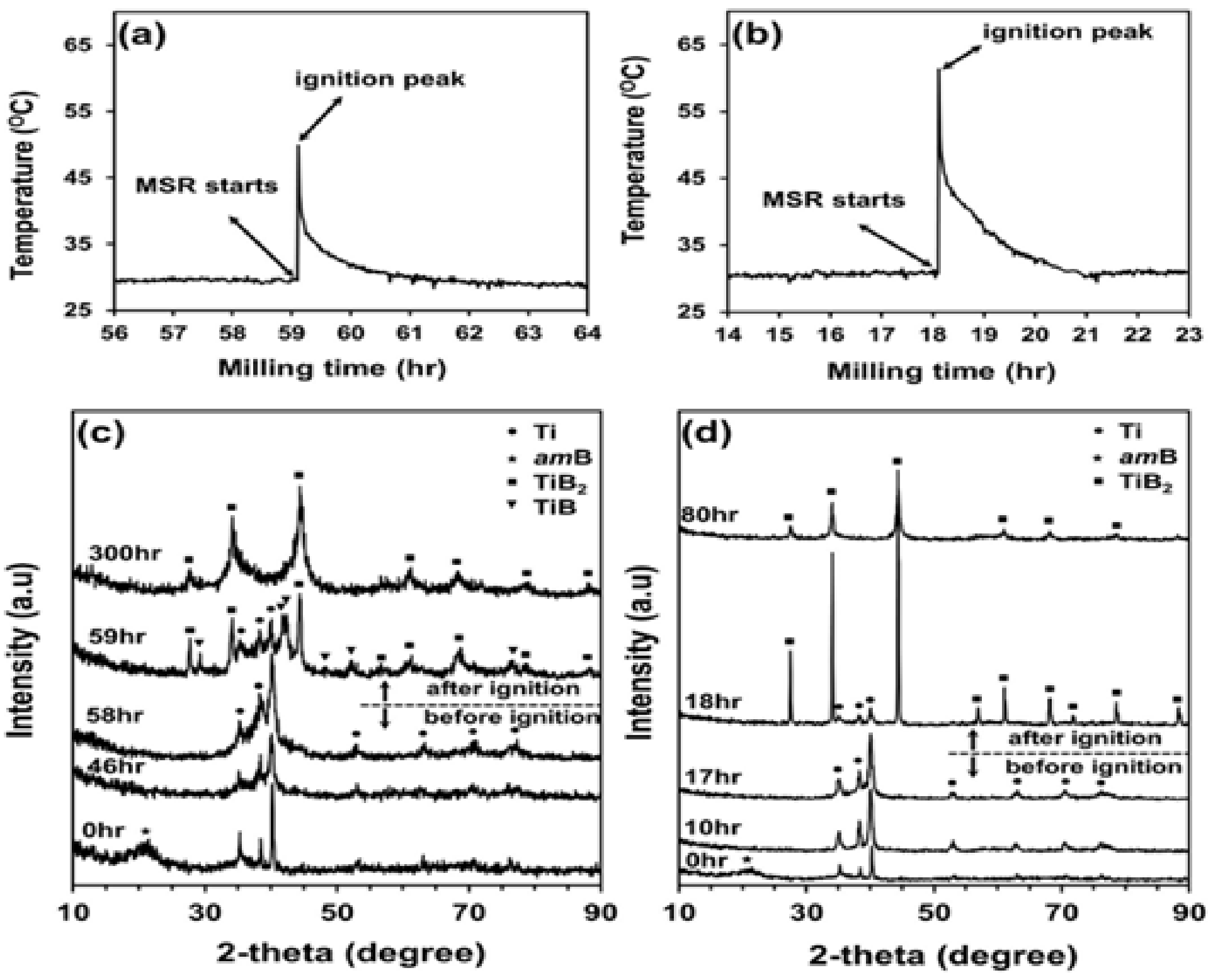

Figure 8 Plots of the external temperature of the vial vs milling time for (a) $Ti_{50}B_{50}$ and (b) $Ti_{34}B_{66}$ mixtures, (c) and (d) show the respective XRD results. Reprinted from [101]. Copyright (2017), with permission from Elsevier.

The XRD pattern of $Ti_{50}B_{50}$ mixture treated for 0 – 58 h of milling did not reveal the formation of new phases, however, a mixture of hexagonal TiB and orthorhombic $TiB_2$ with some residues of non-reacted Ti was observed after 59 h (***Figure 8***c). After 300 hours of milling, the Ti and TiB phases were not identified, leaving $TiB_2$ as the sole thermodynamically stable phase. $TiB_2$ was formed despite the initial composition of Ti:B was 50:50, due to the higher heat formation of $TiB_2$ over TiB. A similar pattern was observed for milled $Ti_{34}B_{66}$ (***Figure 8***). Before the MSR ignition (0-17 hours), Ti peaks exhibited a slight shift to lower Bragg angles. They also exhibited significant broadening, and right after the MSR ignition after 18 hours, distinct peaks associated with $TiB_2$ and minor broader peaks of unreacted Ti were identified. By the 80-hour mark, the $TiB_2$ peaks had notably broadened, and the Ti had disappeared.

Apart from these two case studies, plenty of other publications related to the MSR of borides were published. The basic information regarding these studies is summarized in ***Table 2***.

Table 2. Overview of the studies devoted to MSR investigation of borides published since 2000

| Year | Reagents | Target compounds | Ignition time | Mill type | Ref |
|---|---|---|---|---|---|
| 2019 | $CeCl_3$, $CeO_2$,$B_2O_3$, Mg, $MgH_2$ | $CeB_6$ | 28 min | Mixer mill | [102] |
| 2018 | $B_2O_3$, $HfO_2$, Mg | $HfB_2$ | 12 min | Planetary mill | [103] |
| 2018 | Mg, $Ta_2B_5$, B | $TaB_2$ | 4 min | Planetary mill | [104] |
| 2017 | $Co_3O_4$, $V_2O_5$, $B_2O_3$, Mg | Co-VB | 15 min | Planetary mill | [105] |
| 2017 | Ti, Ta, Nb, V, B | $TiB_2$, VB, $VB_2$, NbB, $NbB_2$ and TaB | TaB (10 min), TiB (7 min), VB (15 and 20 min, depending on the starting composition), NbB (15 and 20 min, depending on the initial composition) | Planetary mill | 106 |
| 2017 | Ti, B | TiB, $TiB_2$ | 59 h | Cylindrical ball mill with external magnet | 101 |
| 2015 | Al, Mg, $B_2O_3$, Nb | $NbB_2$ | 2.4 h | Planetary mill | 107 |
| 2015 | $ZrSiO_4$, Mg, C | $ZrB_2$–SiC–ZrC | 6 min | Planetary mill | 108 |
| 2014 | $TiO_2$, $B_2O_3$, Mg | $TiB_2$ | 14 min | Planetary mill | 109 |
| 2014 | $TiO_2$,$ZrO_2$,$HfO_2$,$HBO_2$ and Al | $TiB_2$-$Al_2O_3$, $ZrB_2$-$Al_2O_3$, $HfB_2$-$Al_2O_3$, (Ti0.5$Zr_{0.5}$)-$Al_2O_3$, ($Ti_{0.5}$Hf0.5)-$Al_2O_3$, ($Zr_{0.5}Hf_{0.5}$)-$Al_2O_3$ | Ranging from 16.5 min to 23.08 min (depending on the initial composition) | Planetary mill | 110 |
| 2014 | $WO_3$, $B_2O_3$, Mg, C | WB, $WB_2$, $W_2B_5$ | 30 min | Planetary mill | 111 |
| 2014 | Mg, $WO_3$, $B_2O_3$ | $W_2B$, W | 20 min | Planetary mill | 112 |
| 2013 | Al, Mg, $ZrO_2$, $B_2O_3$ | $ZrB_2$ | 13 min (reduction by Al) and 6 min (reduction by Mg) | Planetary mill | 113 |
| 2011 | $Al_2O_3$–$TiB_2$ Al, $TiO_2$, $H_3BO_3$ | $TiB_2$ | 1.5 h | Planetary mill | 114 |
| 2011 | Ti, Zr, Hf, B | $Ti_{1-x}Zr_xB_2$, $Ti_{1-x}Hf_xB_2$ | Ranging from 47 to 113 (depending on the initial composition) | Planetary mill | 100 |
| 2006 | Ti, BN, Fe-Cr-Ni (as a binder phase) | TiN/$TiB_2$ | 2 h | Planetary mill | 115 |

### 2.1.3. Silicides

Silicides are also refractory materials widely used as transistors,[116] for lithium-ion battery anodes,[117] or as semiconductors.[118] They were prepared many times by MSR, as the review by Takacs from 2002 already referenced ten studies related to silicides.[21] Among them, the compound that most studies concentrate on is $MoSi_2$.[119–125] As case studies, the reports investigating the effect of partial substitution of silicon by aluminium and that of molybdenum by tungsten on the course of MSR between Mo and Si to yield $MoSi_2$[121,122] were selected. The substitution of silicon by Al takes place according to the equation (2):

$$Mo + 2(1-x)Si + 2xAl \rightarrow Mo(Si_{1-x}Al_x)_2 \quad (2)$$

the ignition time was reduced up to 30% with $x$ being in the range between 0 and 8 at. % [122] (***Figure 9***). This could be explained from the thermodynamic perspective. Al and Si form a eutectic mixture at a relatively low temperature (577 °C), enabling combustion through the formation of a liquid reactant. This liquid reactant facilitates ignition and reduces the ignition time at lower Al concentrations. On the other hand, at higher Al content, the reaction becomes less exothermic, resulting in a longer ignition (***Figure 9***), attributed to the natural properties of Al, particularly its high conductivity, which, while promoting the propagation at the combustion front, also leads to increased heat loss to the environment.

The effect of aluminum, tungsten and combination of both elements on the formation of $MoSi_2$ was reported in ref. [121] For a mere Mo and Si mixture, the ignition was observed at 90 min, indicated by temperature spike from 27.8 °C to 38.9 °C (***Figure 10***I). However, residual molybdenum was still detected from the XRD pattern, suggesting that the MSR did not fully consume the reagents. The addition of tungsten was found to prolong $t_{ig}$, depending on its stoichiometric proportion. With tungsten contents of x= 0.1 and 0.2, MSR still occurred at 90 min, and both $MoSi_2$ and $WSi_2$ phases were identified (***Figure 10***II (b) and (c)). When the tungsten content was increased to x=0.3, $t_{ig}$ was slightly prolonged, but the same phases were observed. At x=0.4, no MSR occurred and the XRD pattern showed unreacted starting materials as the major components, with only traces of product phases even after 150 minutes of milling (***Figure 10***I(e)). In case of Al addition, incorporating Al at concentrations of y=0.05 and 0.1 shortened $t_{ig}$ to 80 min and 70 min, respectively. However, at y=0.2, $t_{ig}$ was the same as $t_{ig}$ when milling only Mo and Si. No MSR detected when Al concentration was increased to y=0.3, as confirmed by distinct XRD patterns showing the presence of both Mo and Si (***Figure 10***III(d)) even after 140 minutes of milling. When Al and W were added simultaneously, MSR was only observed at x=0.1 and y=0.1, with $t_{ig}$ of 100 min. This was further confirmed by the appearance of XRD peaks corresponding to $MoSi_2$, $WSi_2$, $W(Si,Al)_2$, $Mo(Si, Al)_2$ along with residual Mo. In contrast, no MSR was observed at higher concentrations of x=0.2 and y=0.2, or x=0.3 and y=0.3, which is supported by XRD patterns showing only Mo and Si, despite the fact that milling was conducted for 170 and 190 min, respectively.

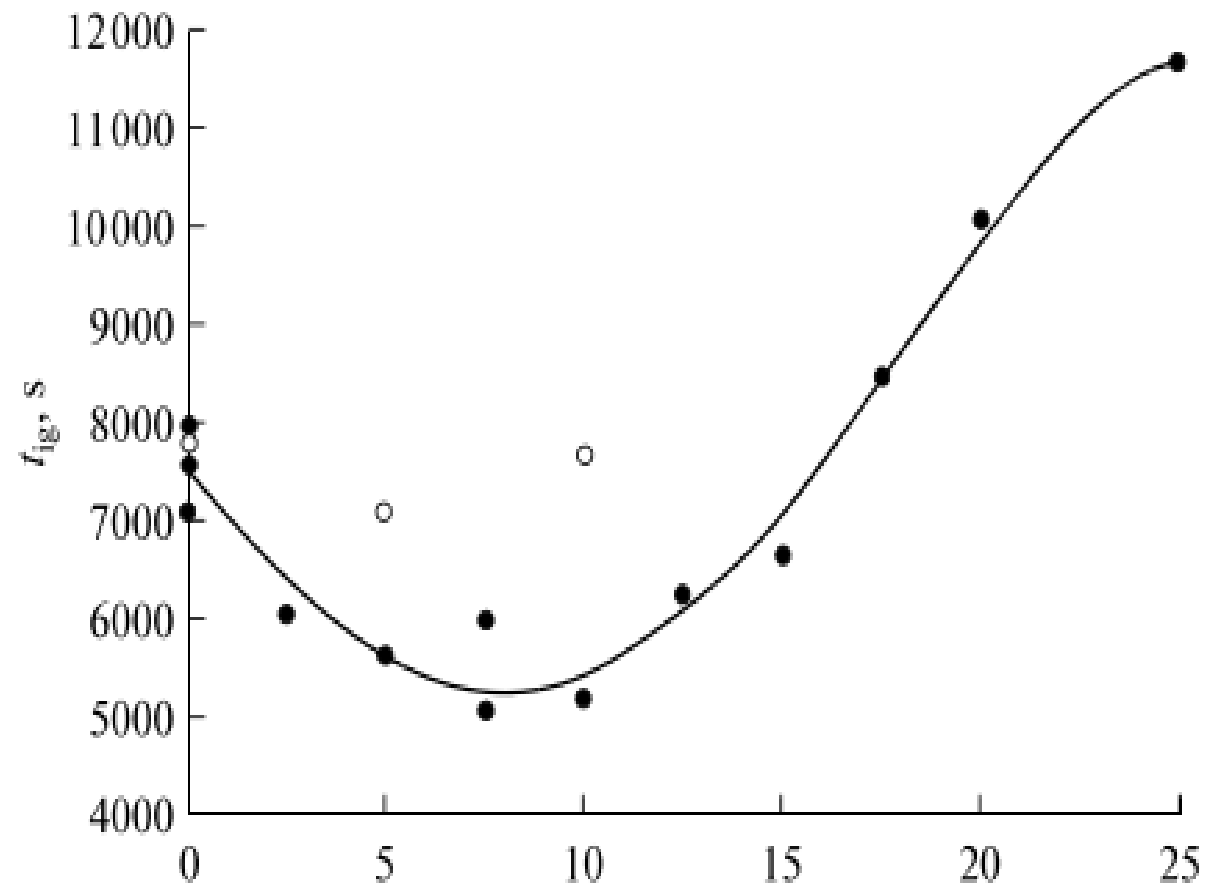

Figure 9 Ignition time $t_{ig}$ as a function of Al content (x). Most data were obtained with 5 g of powder (full symbols.) Some experiments used 8 g powder (open symbols); the ignition time for those was multiplied

by 5/8 to obtain comparable data. Reprinted with permission from Ref. [122]. Copyright, 2013. International Journal of Self-Propagating High-Temperature Synthesis.

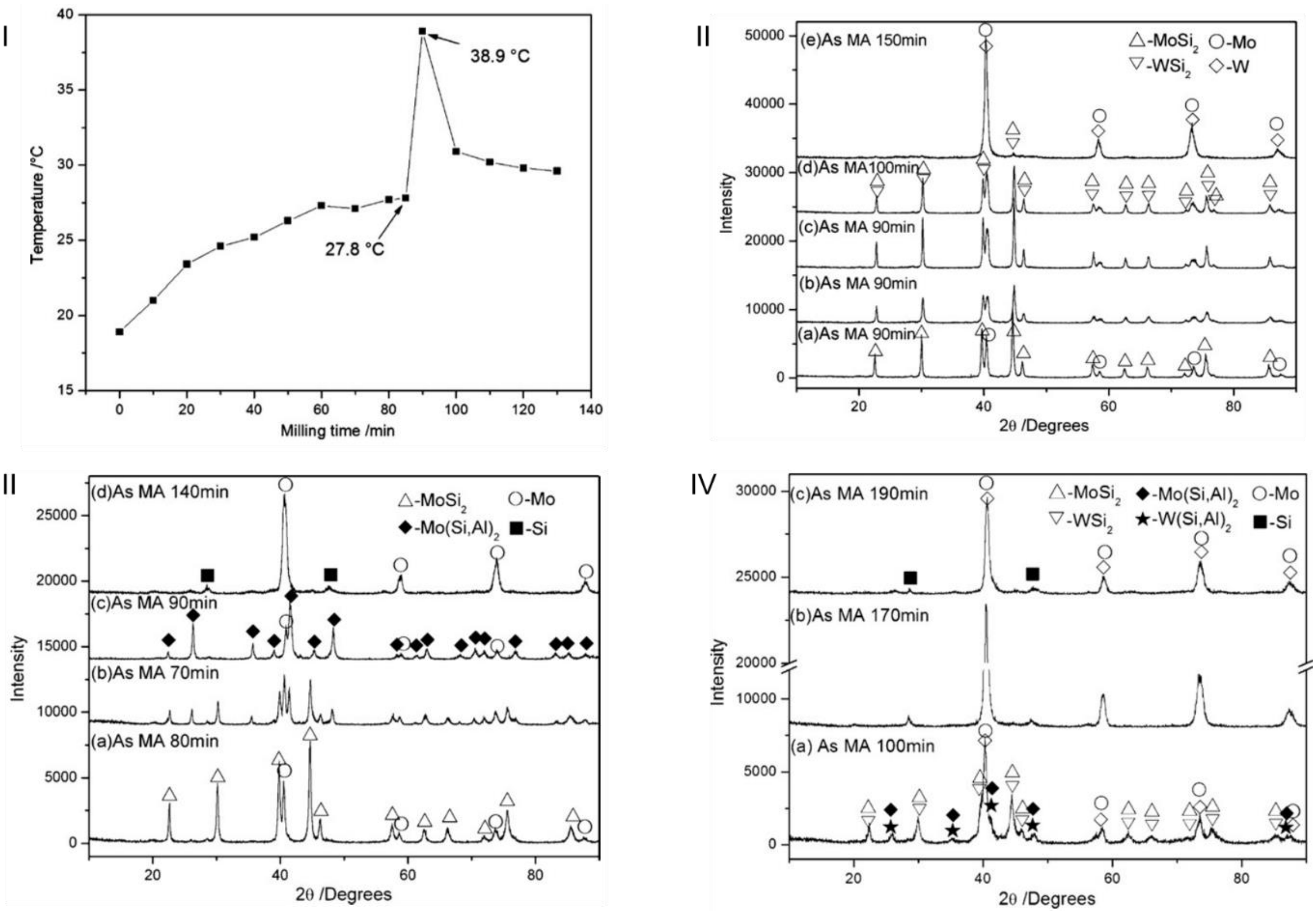

**Figure 10** (I) A temperature trace of the vial during milling of Mo:2Si powders. X-ray diffraction patterns of powder mixture high-energy ball milled for different times (II) W–Mo–Si system: (a) $MoSi_2$ (b) $(W_{0.1},Mo_{0.9})Si_2$; (c) $(W_{0.2},Mo_{0.8})Si_2$; (d) $(W_{0.3},Mo_{0.7})Si_2$; (e) $(W_{0.4},Mo_{0.6})Si_2$.(III) Mo–Si-Al system: (a) $Mo(Si_{0.95},Al_{0.05})_2$; (b) $Mo(Si_{0.9},Al_{0.1})_2$; (c) $Mo(Si_{0.8},Al_{0.2})_2$; (d) $Mo(Si_{0.7},Al_{0.3})_2$, (IV) W–Mo–Si–Al system: $(W_{0.1},Mo_{0.9})(Si_{0.9},Al_{0.1})_2$; (b) $(W_{0.2},Mo_{0.8})(Si_{0.8},Al_{0.2})_2$; (c) $(W_{0.3},Mo_{0.7})(Si_{0.7},Al_{0.3})_2$. Reprinted from [121], Copyright (2010), with permission from Elsevier.

*Table 3* provides a list of relevant studies related to the preparation of silicides via MSR over the specified timespan.

Table 3. Overview of the studies devoted to MSR investigation of silicides since 2002

| Year | Reagents | Target compounds | Ignition time | Mill type | Ref. |
|---|---|---|---|---|---|
| 2023 | Ti, Si | $Ti_5Si_3$ | 4 min to 30 min, depending on the rotation speed | Planetary mill | [126] |
| 2019 | $WO_3$, Si, Mg, $SiO_2$, W | $WSi_2$,$W_5Si_3$, W | 1 min | Mixer Mill | [127] |
| 2017 | $MoO_3$, Mo, Si and Al | $Mo_5Si_3$, $Mo_3Si$ | 1 h | Planetary mill | [128] |
| 2016 | $MoO_3$, Ni, Si, Al | $Mo_2Ni_3Si$ | 10 h | Planetary mill | [129] |
| 2014 | Ce,Si | CeSi- various stoichiometries ($Ce_5Si_3$,$Ce_3Si_2$, CeSi, $CeSi_{1.8}$, $Ce_5Si_4$) | Ranging from 35 to 68 minutes, depends on milling speed | Planetary mill | 130 |
| 2014 | $MoO_3$, Si, Al | $MoSi_2$-mullite | 23-30 min | Planetary mill | 131 |
| 2013 | Mo, Si, Al | $MoSi_2$, $Mo(SiAl)_2$ | Ranging from 5000 s to 12000 s, depends on the content of Al | Planetary mill | 122 |
| 2013 | Mo, Si | Mo-Si | - | Planetary mill | 132 |
| 2013 | Mo, Si | $MoSi_2$-$Si_3N_4$ | 5 h | Planetary mill | 133 |
| 2013 | $MoO_3$, Si, Al | $MoSi_2$ | - | Planetary mill | 120 |
| 2011 | Ti, Si, Ge | $Ti_5Si_3$-$Ti_5Ge_3$ | Ranging to 19 to 40 min (depends on starting compositions) | Planetary mill | 134 |
| 2011 | Mo, Si | $MoSi_2$ | Between 105 and 110 min | Planetary mill | 135 |
| 2010 | Mo, Si, W, Al | $MoSi_2$ | 90 min to 100 min (addition of W), and ranging from 80 min to 140 min (addition of Al) | Planetary mill | 121 |
| 2008 | Ti, Si, $N_2$ gas | Ti-Si intermetallic | 10, 12, and 19 min | Planetary mill | 119 |
| 2008 | Mo, Si | MoSi | 60 to 130 min based on molar ratio of Mo:Si | Planetary mill | 136 |
| 2006 | $MoO_3$, $SiO_2$, Al | α-$MoSi_2$ and β-$MoSi_2$ | 3 h | Planetary mill | 124 |
| 2004 | Ti, Al, Si, C | TiC, $Ti_5Si_3$ | 5 h | Planetary mill | 96 |
| 2004 | Mo, Si | $MoSi_2$ | 8 h | Planetary mill | 123 |
| 2004 | Ti, $Si_3N_4$ | TiN and $Ti_5Si_3$ | 100 min | Mixer mill | 42 |

### 2.1.4. Carbonitrides

The category of carbonitrides, specifically those comprising a Ti(C,N) solid solution, is classified as cermets, frequently complemented with a metallic binder such as Co or Ni to enhance toughness and durability.[137,138] These composite materials are predominantly utilized in cutting tool applications, capitalizing on their superior mechanical properties- particularly outstanding wear resistance and stability under elevated thermal conditions.[139,140]

The mechanochemical synthesis of carbonitrides usually involves introducing nitrogen gas as the nitrogen source into the milling vial where metal and carbon are being treated, allowing controlled nitrogen incorporation into the matrix to achieve the desired stoichiometry and phase distribution. Before a significant effort to prepare carbonitrides via MSR started, nitrides were prepared in this way. Namely in 2002, titanium nitride was successfully synthesized by Gotor et al., igniting a MSR when milling titanium metal in nitrogen atmosphere.[141] No work devoted to nitrides or carbonitrides synthesis via MSR was reported in the review by Takacs[21]. The majority of works related to carbonitride preparation via MSR is related to titanium-based compounds, so we have selected three case studies from this system.

Cordoba et al.[142] synthesized complex titanium niobium carbonitrides, $Ti_yNb_{1-y}C_xN_{1-x}$, by manipulating the ratios of Ti, Nb, and C in two specific formulations Ti : Nb : C = 1 : 1 : 1 (sample A) and 3 : 1 : 2 (sample B). The ignition times, indicative of reaction initiation, were observed at 58 minutes for sample A and 49 minutes for sample B. Milling yielded nanometric powders characterized by a homogeneous chemical composition and monophasic carbonitride structure. Sample A, which possessed a composition of $Ti_{0.59}Nb_{0.41}C_{0.68}N_{0.32}$, demonstrated a smaller crystallite size and greater microstrain than sample B, suggesting enhanced lattice distortion likely due to its more balanced Ti and Nb content in comparison with sample B. However, they do not explain why the product stoichiometry is significantly different from the starting one. In contrast, sample B, with a composition of $Ti_{0.79}Nb_{0.21}C_{0.66}N_{0.34}$, exhibited a larger coherent diffracting domain, attributed to its increased Ti content. The higher Ti concentration in sample B also led to a significantly elevated adiabatic temperature, pointing to a more exothermic reaction mechanism and accelerated phase formation. The same complex carbonitride powders were also synthesized by Borrell et al. in 2012.[143] The synthesis of $TiC_xN_{1-x}$ via MSR has been shown to significantly enhance the physical properties of the material, particularly in terms of hardness and toughness. Using activated carbon as the carbon source resulted in products with a higher specific surface area and smaller diffraction domain sizes compared to those synthesized with graphite. This difference is attributed to the increased reactivity of activated carbon, which promotes finer particle distribution. On the other hand, graphite offers a unique advantage due to its ability to coat titanium particles, facilitating the initiation of the combustion reaction under lower-energy milling conditions.[34] A subsequent study by the same research group [144] utilized carbon nanofibers (CNFs) instead of activated carbon as the carbon source. Before the actual reaction with titanium and nitrogen, the as-received CNFs were mechanically activated to obtain milled CNFs. The particle size of CNFs was decreased to the nanometric scale compared to the micrometric scale of the starting material. The ignition time of the subsequent MSR was significantly reduced when using milled CNFs compared with both graphite and non-milled CNFs. The ignition time for the reaction was approximately 45 minutes with non-milled CNFs, but this dropped to just 20 minutes when milled CNFs were used. This decrease in the ignition time is attributed to the reduced particle size of CNFs during prior mechanical activation. The reduction of CNF particle size to the nanometric scale not only facilitates faster combustion initiation but also improves the uniformity of the carbon distribution within the reaction matrix. This structural refinement enables a more rapid and efficient synthesis process, demonstrating the critical role of particle size in modulating reactivity and reaction kinetics in MSR.

The relevant works on MSRs of carbonitrides are summarized in ***Table 4***.

Table 4. Overview of the studies devoted to MSR investigation of carbonitrides since 2002

| Year | Reagents | Target compounds | Ignition time | Mill type | Ref. |
|---|---|---|---|---|---|
| 2023 | Hf, Zr, C, N | (Hf,Zr)(C,N) solid solution | 10 min | Planetary mill | 145 |
| 2018 | Mg, $Nb_2O_5$, $C_3H_6N_6$ | Nb(C,N) | 5 min | Planetary mill | 146 |
| 2018 | TiN, TiC, $Si_3N_4$ and SiC | TiC-N | 50 to 80, depend on TiN content | Planetary mill | 147 |
| 2017 | Mg, $B_2O_3$, $C_3H_6N_6$ | BCN | 30, 35 and 50 min, depend of melamine content | Planetary mill | 148 |
| 2015 | V + C under $N_2$ atmosphere | V-C-N | ranging from 32 to 61 min, depend on C-V powder mixtures | Planetary mill | 149 |
| 2012 | Ti, Ta, C, N, Co | $Ti_{0.8}Ta_{0.2}C_{0.5}N_{0.5}$— 20 wt.% Co | 45.1 min | Planetary mill | 138 |
| 2012 | Ti, C, $N_2$, Co | $TiC_xN_{1-x}$-15%Co | 15 mins | Planetary mill | 144 |
| 2012 | Ti, Ta, C, Co | $Ti_xTa_{1-x}C_yN_{1-y}$ | 37 min | Planetary mill | 137 |
| 2012 | Ti, Nb, C | (Ti,Ta,Nb)$C_xN_{1-x}$ | ranging from 43 to 53 min, depends on the initial powder mixture | Planetary mill | 150 |
| 2012 | Ti, Nb, C | $Ti_yNb_{1-y}C_xN_{1-x}$ | Ranging from 39 to 51 depends on initial powder mixture | Planetary mill | 143 |
| 2011 | Ti, C(black carbon) | $TiC_7N_3$, $TiC_5N_5$, and $TiC_3N_7$ | 24 h | Planetary mill | 151 |
| 2011 | Ti, $TiO_2$, Al, C | Ti-C-N,$Al_2O_3$ | 14 h | Planetary mill | 152 |
| 2010 | Ti, C, $N_2$ gas | $TiC_xN_{1-x}$ | 45 min for original CNFs and 25 min for milled CNFs | Planetary mill | 144 |
| 2009 | Ti, Zr, V, Ta, Hf, C, $N_2$ | $Ti_yMT_{1-y}C_xN_{1-x}$ (MT: Zr, V, Ta, Hf) | ranging from 31 to 60 min | Planetary mill | 153 |
| 2008 | Ti, C, N, W, WC, $Mo_2C$ | $TiC_xN_{1-x}$ | ranging from 24 to 48 min | Planetary mill | 154 |
| 2006 | Ti, Nb, C | $Ti_yNb_{1-y}C_xN_{1-x}$ (TiC, TiN, NbC, NbN) | 58 and 49 min | Planetary mill | 142 |
| 2005 | Ti, C (activated carbon or graphite) | $TiC_xN_{1-x}$ | 20-43 min (activated carbon as C source) 50 min (graphite) | Planetary mill | 34 |

## 2.2. Metal chalcogenides

Metal chalcogenides have garnered significant attention in various fields, such as catalysis, thermoelectrics, energy storage, and biomedicine, due to their exceptional physicochemical properties.[155,156] Mechanochemical synthesis is a well-established method for the preparation of metal chalcogenides, utilizing elements or compounds as reagents.[157] When using elements, MSR pathway is possible.[21]

The research dedicated to MSRs in metal chalcogenides is notably less extensive compared to the more thoroughly explored metalloids, albeit it was the first system where MSR was reported.[24] In the review by Takacs,[21] 27 studies related to metal chalcogenides synthesized through the MSR are mentioned. Also, he was personally engaged in researching this topic (e.g. ).[158] The pioneering investigation of MSRs in metal chalcogenides involved the reactions of some four selected metals (Cd, IN, Sn, Zn) with either sulfur, selenium or tellurium.[24] The MSRs were observed within minutes to hours, depending on the system. The hypothesis that the MSR ignition time is related to the hardness of the metal and chalcogen was formulated in this study. Due to the high value of this study, the table mentioning various ignition times for different systems presented there is also reprinted here as ***Table 5***. The table also shows the ΔH/C ratios, temperature difference detected and the hardness data.

**Table 5.** Information about MSRs for metal-chalcogen reactions investigated by Tschakarov and Rusanov in:[24] *ΔH/C* ratios, the ignition time ($t_{ig}$), and temperature difference during MSR (*ΔT*) and hardness data

| Reaction | *ΔH/C* (K) | $t_{ig}$ (min) | *ΔT* (K) | Microhardness | |
|---|---|---|---|---|---|
| | | | | Metal | Chalcogen |
| Zn + S | 4520 | 365 | 30 | 530 | 290 |
| Zn + Se | 3140 | 312 | 23 | 530 | 450 |
| Zn + Te | 2400 | 300 | 2 | 530 | 500 |
| | | | | | |
| Cd + S | 3070 | 78 | 17 | 290 | 290 |
| Cd + Se | 2920 | 46 | 7 | 290 | 450 |
| Cd + Te | 2020 | 55 | 3 | 290 | 500 |
| | | | | | |
| 2In + 3S | 3010 | 25 | 6 | 100 | 290 |
| 2In + 3Se | 2310 | 20 | 3 | 100 | 450 |
| 2In + 3Te | 1550 | 12 | 2 | 100 | 500 |
| | | | | | |
| Sn + S | 2190 | 110 | 2 | 90 | 290 |
| Sn + Se | 1790 | 42 | 2 | 90 | 450 |
| Sn + Te | 1170 | 22 | 2 | 90 | 500 |
| Sn + 2S | 2190 | 258 | 3 | 90 | 290 |
| Sn + 2Se | 1740 | 21 | 3 | 90 | 450 |
| Sn + 2Te | ? | 42 | 1 | 90 | 500 |
| | | | | | |
| Pb + S | 1980 | 40 | 3 | - | 290 |
| Pb + Se | 1990 | 12 | 3 | - | 450 |
| Pb + Te | 1360 | 14 | 3 | - | 500 |

**Table 5** presents data on the hardness of various metals and chalcogens, revealing a linear correlation between metal hardness and ignition time in MSRs. To demonstrate this, a figure showing the relationship between the hardness of the metals and ignition time for corresponding sulfides and selenides in the mentioned study is given in ***Figure 11***. It was found that softer metals and harder chalcogens tend to exhibit shorter ignition times, suggesting a relationship between material properties and reaction kinetics.[24] However, recent findings by Baláž et al.[45] challenge the validity of this correlation, indicating that additional factors may influence the ignition behaviour, for example, the particle size. To visualize this, the same dependence as in ***Figure 11***a with the same systems, showing completely different behaviour, is shown as ***Figure 11***b . Additionally, nickel-based chalcogenides were also investigated and as it is shown in ***Figure 11***b, sulfides formed more easily than selenides, being in contradiction to what was previously observed. Research aiming to find an explanation for this discrepancy is being conducted these days. As will be shown later on in the review, also other researchers propose different hypotheses which is in contradiction with the early claims made by Bulgarian researchers. Nevertheless, their discoveries sparked a new research pathway within mechanochemistry.

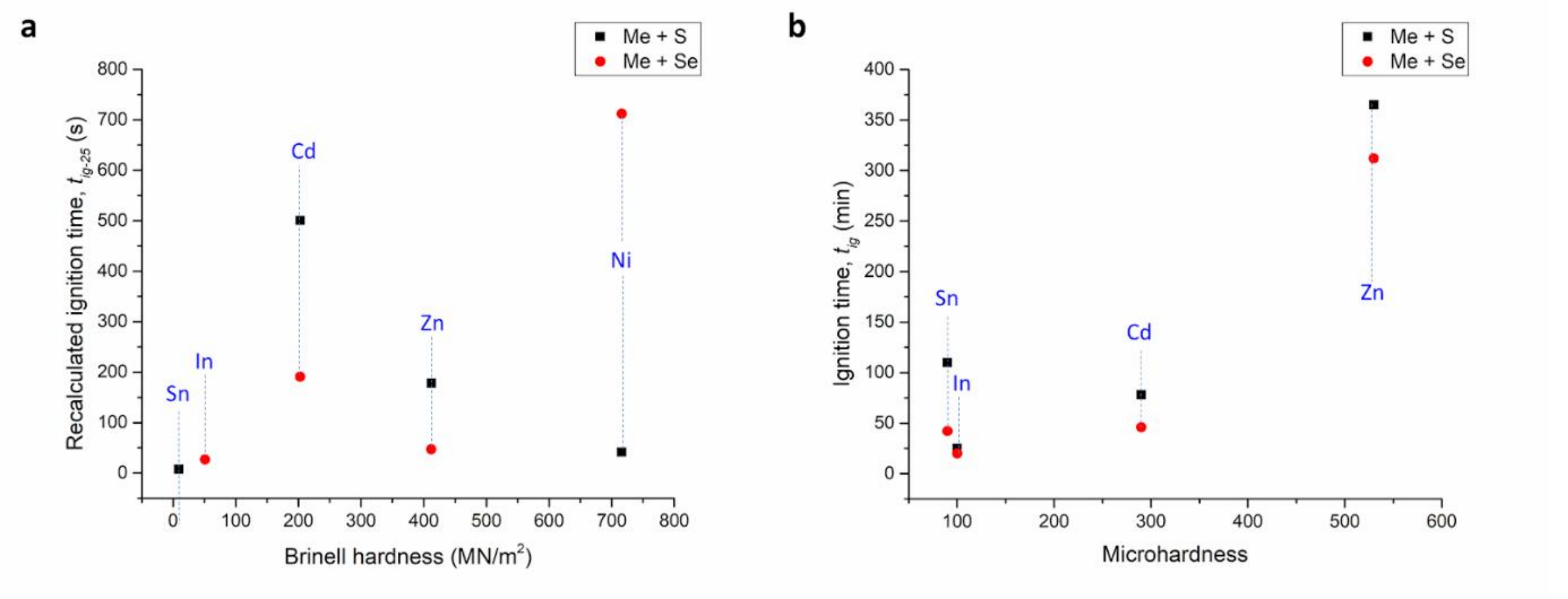

**Figure 11** Influence of hardness on time of ignition: (a) data from[24] Reprinted from ref. [24] Copyright 1982, with permission from Elsevier." (b) data from[45]. 2024. Licensed under CC BY-NC.

As this chapter deals with the group of compounds that our research group has a large experience in, we have selected two studies from our lab as case studies.

In 2016, we published a study reporting the occurrence of MSR event for the copper-sulfur (not respecting a rule-of-thumb exothermicity rule) system for the first time. The MSR event is referred to as "explosive" in the mentioned study, due to lack of knowledge back then. It turned out that the properties of the metallic reagents have a decisive effect on the potential occurrence of MSR. Copper and sulfur powders from four and two different manufacturers, respectively, were used for the experiments. In ***Figure 12***a-d, the temperature and pressure monitoring curves can be seen. Whereas in two cases, an abrupt pressure increase (proof of MSR) was observed after a few seconds of treatment, a gradual pressure increase was observed within 5 minutes in two other cases. The apparent drop of temperature at the moment of MSR is an artifact. Due to the placement of the temperature sensor, the used experimental setup does not allow for the detection of a temperature spike at the moment of MSR, as it takes time until the temperature increase gets recognized by the temperature sensor. In the case of pressure, it is immediately detected. The detailed analysis has shown that the copper powders used for the reaction where the MSR occurred were fine and possessed flake- and needle-like morphology (Cu(M)). The XRD patterns (***Figure 12***e) confirmed the successful rapid formation of copper sulfides in these cases, whereas only traces of products and large amounts of elemental reagents were evidenced after 5 minutes of treatment when utilizing copper powders with spherical particles (Cu(Pa)). As products, always a mixture of covellite (CuS) and digenite ($Cu_9S_5$) was formed, depending on the used sulfur source. Digenite was found to be present either as a hexagonal (hex.) or cubic (cub.) phase.

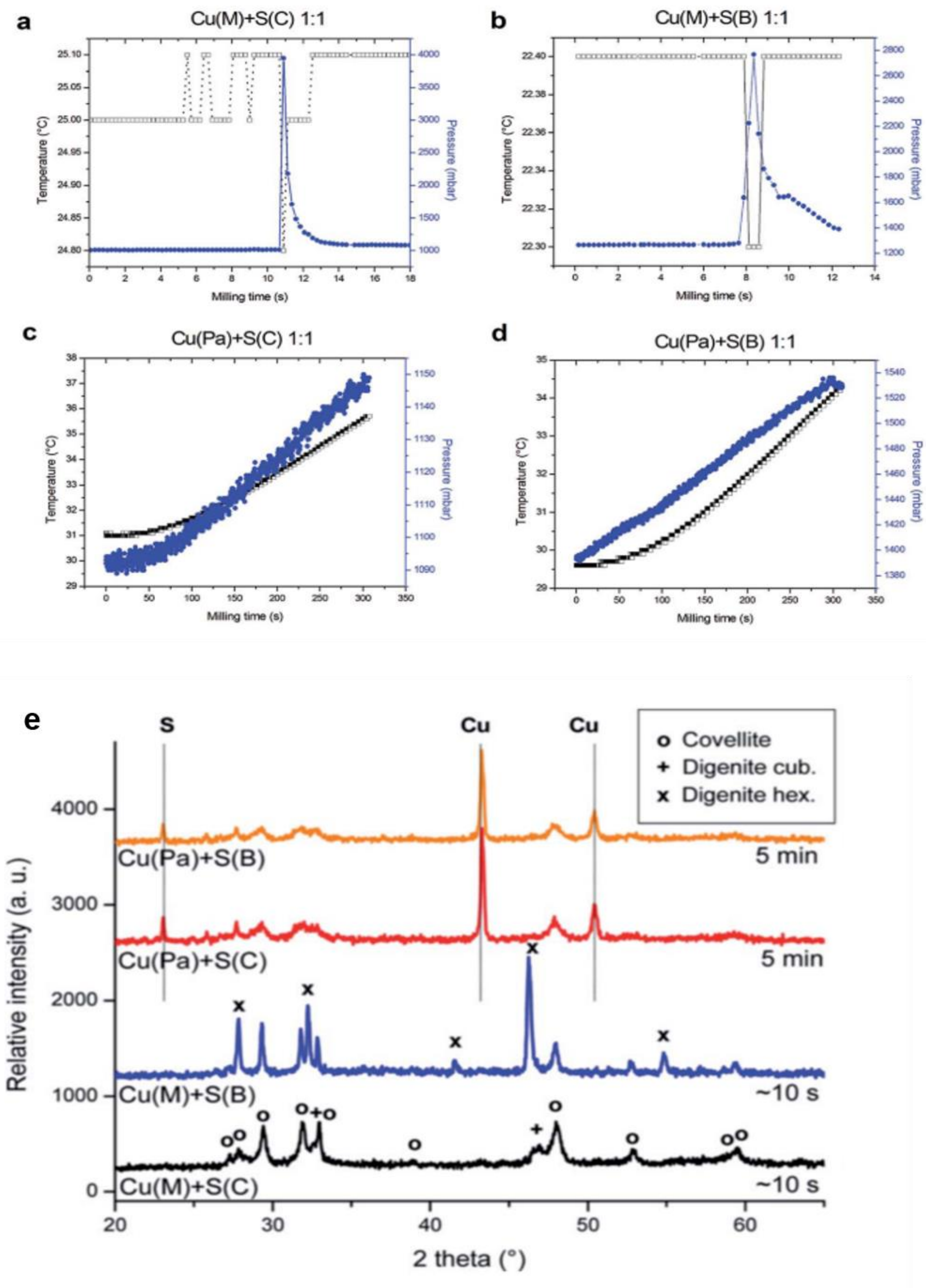

**Figure 12** Gas pressure and temperature changes during milling. Cu(M)—electrolytically prepared copper with flaky and needle-like morphology, Cu(Pa)—copper prepared in the form of atomized powder with spherical morphology, S(B), S(C)—sulphur from different producers. Time of milling: (a, b) until MSR (approximately 10 seconds), (c, d) 5 minutes, (e) XRD patterns of the milled Cu:S 1:1 mixtures. The time of milling is given in the figure. Reprinted with permission from Ref. [36]. Copyright 2016, Royal Society Chemistry.

Recently, in the study being part of the first issue of the journal RSC Mechanochemistry, we re-visited quite similar set of mechanochemical syntheses of metal chalcogenides via MSR as in the pioneering study by Tschakarov and Rusanov.[24] We managed to synthesize some binary metal chalcogenides (CdS, CdSe, $In_2S_3$, NiS, NiSe, PbS, PbSe SnSe, ZnS and ZnSe) in less than 10 minutes. This work reports the MSR occurrence in the Ni-based systems for the first time. The study also contains considerations about different factors affecting the MSRs, and it seems that the most important role is played by the particle size of the reagents and bulk modulus. The gas pressure monitoring curves obtained during milling for different systems are shown in ***Figure 13***.

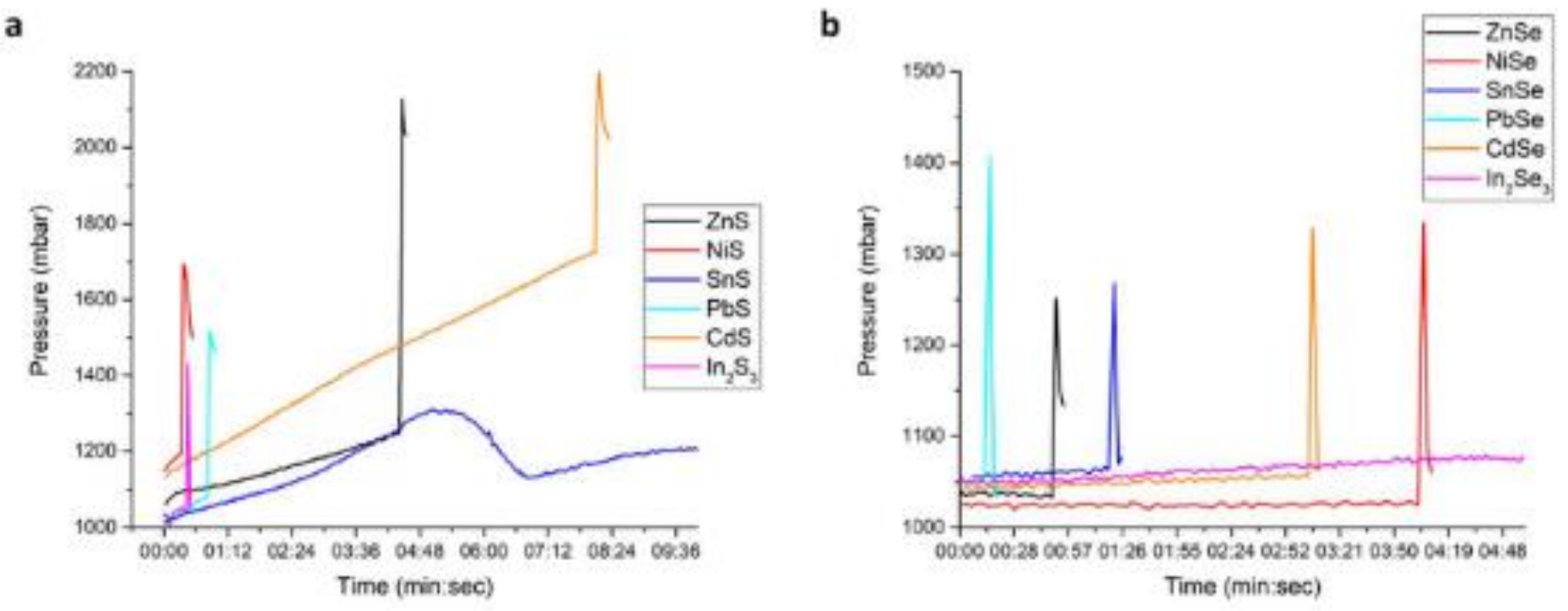

**Figure 13** Gas pressure changes during milling (representative curves): (a) Me–S mixtures treated at 750 rpm and (b) Me–Se mixtures treated at 400 rpm in a planetary ball mill. Selenides were too reactive at 750 rpm, so the reactions needed to be slowed down. [45]. 2024. Licensed under CC BY-NC.

From our point of view, one more study deserves a special paragraph in this review due to its important outcome for the fundamental understanding of MSRs. It involves the mechanochemical synthesis of metal selenides. Ohtani et al. [159] studied the effect of pre-milling in the mechanochemical synthesis of CdSe and ZnSe. Without pre-milling, the ignition times ($t_{ig}$) for CdSe and ZnSe were observed to be 2 minutes and 8 minutes, respectively. However, when each constituent was pre-milled for 15 minutes, the formation of both CdSe and ZnSe occurred very quickly, within just 5 seconds of milling. This suggests that during the activation period, the mechanical energy supplied by ball milling is primarily used to activate elemental particles rather than to drive for the reaction itself [35].

The summary of the studies devoted to the investigation of MSRs in the preparation of metal chalcogenides published since 2002 is presented in ***Table 6***.

Table 6. Overview of the studies devoted to MSR investigation of metal chalcogenides since 2002

| Year | Reagents | Target compounds | Ignition time | Mill type | Ref. |
|---|---|---|---|---|---|
| 2026 | Cu, S | CuS | up to 2 min | Planetary mill (ball-free jar) | 160 |
| 2024 | Zn, Pb, Ni, In, Sn, Cd, Se | ZnSe, PbSe, NiSe, $In_2Se_3$, SnSe, CdSe | 62 s (ZnSe), 13 s (PbSe), 195 s (NiSe), No MSR ($In_2Se_3$), 85 s (SnSe), 188 s (CdSe) | Planetary mill | 45 |
| 2024 | Zn, Pb, Ni, In, Sn, Cd, Se | ZnS, PbS, NiS, $In_2S_3$, SnS, CdS | 211 s (ZnS), 55 s (PbS), 11 s (NiS), 27 S ($In_2S_3$), No MSR (SnS), 496 s (CdS) | Planetary mill | 45 |
| 2021 | Zn, S, Se, Mn | $Mn^{2+}$-doped $ZnS_xSe_{1-x}$ | Ranging from 8 min to 80 min depends on Se content | Planetary mill | 161 |
| 2021 | Zn, S | ZnS | 30 min at 27 Hz 20 min at 30 Hz | Shaker mill | 38 |
| 2020 | Mg, $MgH_2$, Cr, S | $MgCr_2S_4$ | 8 h | Mixer mill | 162 |
| 2020 | Fe, Ni, S | $Fe_{4.5}Ni_{4.5}S_8$ | Ranging from 8 min to 15 min (depending on BPR) | Planetary mill | 44 |
| 2019 | Zn, S, Se | $ZnS_xSe_{1-x}$ | Ranging from 8 min to 80 min depends on Se content | Planetary mill | 163 |
| 2017 | Cu, S | CuS and $Cu_2S$ | 10 s | Planetary mill | 164 |
| 2016 | Cu, S | CuS and $Cu_2S$ | 10 s | Planetary mill | 36 |
| 2015 | Cd, Se, Zn | CdSe, ZnSe | CdSe: 5 s ZnSe: 8 min | Planetary mill | 159 |
| 2014 | Cu, In, S | $CuInS_2$ | 120 min | Planetary mill | 165 |
| 2014 | Ce, S | CeS | 1 h | Planetary mill | 166 |
| 2013 | Zn, Se | ZnSe | 14 min | Planetary mill | 35 |
| 2012 | $SrSO_4$, Mg | SrS | 5 min | Planetary mill | 167 |
| 2011 | $SrSO_4$, Al | SrS | 35 min | Planetary mill | 168 |
| 2011 | Cu, In, Ga, Se | $CuIn_{0.75}Ga_{0.25}Se_2$ | 45 min | Mixer mill | 169 |
| 2010 | Cu, In, Se | $CuInSe_2$ | 45 min | Planetary mill | 165 |

### 2.3. Metal oxides

Thermite reactions, which involve the reaction between a metal and metallic or non-metallic oxides [170], are characterized by their highly exothermic and self-sustaining nature [171]. The first thermite reaction was discovered in 1893 by German chemist Goldschmidt [172]. Since this discovery, numerous studies related to displacement reactions involving oxygen have been conducted. These reactions can also be performed as MSRs by employing a displacement reaction, represented by the following equation:

$$A_XO + yB \rightarrow xA + B_yO \quad (3)$$

In the pioneering review by Takacs was published in 2002,[21] twenty-two reactions involving oxides were already mentioned. The research in the field continued to evolve during the following years. Aluminum and magnesium are the mostly used reduction agents in the recent studies, [173–175] leading to what are known as aluminothermic and magnesiothermic reactions, respectively. For this chapter, we have selected three case studies utilizing nickel oxide.

A displacement reaction between nickel oxide and magnesium/titanium in the preparation of metallic nickel was performed via the reduction process in.[173] The study was conducted in a planetary ball mill under an argon atmosphere to prevent oxidation, following the equations:

$$NiO + Mg \rightarrow Ni + MgO \qquad T_{ad} = 5500\ \mathrm{K} \quad (4)$$

$$2NiO + Ti \rightarrow 2Ni + TiO_2 \qquad T_{ad} = 4000\ \mathrm{K} \quad (5)$$

Adjusting the composition closer to stoichiometric ratios shifted the reaction mechanism from a gradual progression to a self-sustaining combustion due to the systems' high exothermicity, as indicated by the calculated adiabatic temperature $T_{ad}$. XRD analysis was used to assess the phase composition present after 30 hours of milling when changing the amounts of reducing agents for both systems, as illustrated in ***Figure 14***a a and b for the NiO-Mg and the NiO-Ti systems, respectively. Before milling (bottommost XRD patterns), the diffraction peaks corresponded to the initial phases of NiO and Mg/Ti. After 30 hours of milling, with incremental increases in Mg content, the diffraction patterns displayed peak shifts to lower 2θ angles with the increase in Mg content ($x_{Mg}$). This shift was attributed to atomic disorder induced by mechanical processing. The presence of small amount of elemental nickel can be clearly observed from $x_{Mg}$ being at least 30%. However, the clear formation of MgO and high amount of Ni with mutual disappearance of the diffraction peaks of elemental reagents can be only confirmed when $x_{Mg}$ = 50. The MSR ignition time was affected by the amount of Mg. The reduction of NiO into Ni began when $x_{Mg}$=20; however, a gradual transformation was observed. The MSR occurred when the $x_{Mg}$>40, resulting in an immediate 90% of reduction. Thus, the MSR can be confirmed only in this case. A similar result was observed in the NiO-Ti system (***Figure 14***b). As more Ti was mixed with NiO, the Ti phases at 2θ (112) and (201) disappeared, while the intensity of Ni (111) became more prominent. This indicates a phase transformation and atomic reordering, consistent with the increased incorporation of Ti. The changes in the diffraction pattern further corroborate the influence of Ti content on the structural evolution of the system during the mechanochemical reaction. The MSR was observed when $x_{Ti}$ was at least equal to 25 (***Figure 14***b).

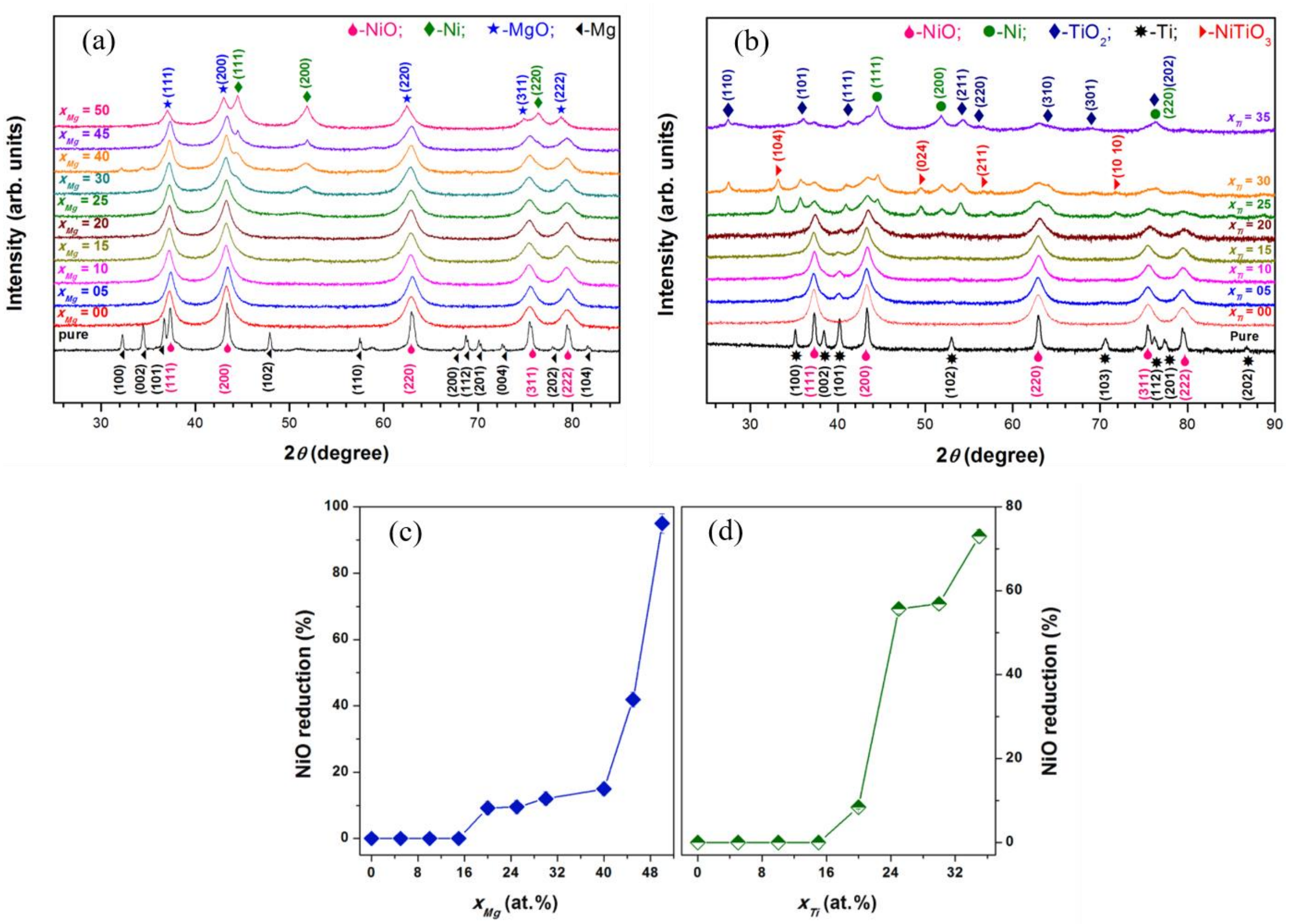

Figure 14 XRD patterns of (a) pure as-mixed and 30 h milled NiO-Mg mixtures with $x_{Mg}$=0–50, (b) pure as-mixed and 30 h milled NiO-Ti mixtures with $x_{Ti}$ =0–35. The degree of NiO reduction as a function of (c) $x_{Mg}$ and (d) $x_{Ti}$. Reprinted from Ref. [173] Copyright (2020), with permission from Elsevier.

As shown in ***Figure 14***c and d, increasing the amount of reducing agents enhances the percentage of NiO reduction, albeit some critical value is required. The high degree of reduction of NiO observed at $x_{Mg}$=50% and xTi=25%. Under these conditions, MSR occurred, so achieving a MSR requires reaching a critical concentration of the reducing agent.

The aluminothermic reaction is a well-established method for reducing metal oxides, leveraging the substantial heat generated during milling. This heat can lead to the combustion of aluminum in the presence of air, resulting in the formation of aluminum oxide through an exothermic reaction. In this case, also the NiAl-$Al_2O_3$ alloy can be obtained. The intense heat release often allows the reduced metal to be obtained in a molten state, enhancing the efficiency of the aluminothermic process for metal oxide synthesis. Synthesis of the complex NiAl/$Al_2O_3$ composite in the presence of stearic acid as process controlling agent (PCA) to avoid oxidation was studied in.[174] The process follows the equation:

$$13Al + 8Ni + 3NiO \rightarrow NiAl + Al_2O_3 \quad (6)$$

The MSR was ignited after 110 minutes and the crystallite size of the obtained intermetallic NiAl and $Al_2O_3$ was 34 and 53 nm, respectively.

Apart from the two case studies describe above, four other reports related to MSR in oxide systems were found. Namely, the synthesis of $MgTiO_3$,[176] Fe-$Y_2O_3$,[177] NiAl-$Al_2O_3$[174] and Fe-$Al_2O_3$[178] was successfully performed utilizing MSR.

| Year | Reagents | Target compounds | Ignition time | Mill type | Ref. |
|---|---|---|---|---|---|
| 2025 | $V_2O_5$, Mg | $MgV_2O_4$ | 20 min | planetary | 179 |
| 2020 | NiO, Mg, Ti | Ni | Not mentioned | Planetary mill | 173 |
| 2018 | Ni, NiO, Al, stearic acid | NiAl-$Al_2O_3$ | 10 min | Planetary mill | 174 |
| 2018 | $CaWO_4$ (from $Na_2WO_4.2H_2O$ and $CaCl_2.2H_2O$), Mg, Al | W | 20-30 min | planetary | 65 |
| 2014 | Mg, MgO, $TiO_2$, HA | $MgTiO_3$ | 4 h and 8 h | Planetary mill | 176 |
| 2011 | $Y_3Fe$, $Fe_2O_3$ | Fe-$Y_2O_3$ | 24 h | Vibratory mill | 177 |
| 2004 | $Fe_2O_3$, Al | Fe-$Al_2O_3$ | 10 h | Planetary mill | 178 |

## 2.4. Organic systems

The investigation of MSRs for organic reactions is quite rare, and only a few reports are available. As the two key studies, the synthesis of porous carbon materials performed by the Borchardt group was selected.[37,53] Specifically, $sp^2$-bonded porous carbon was prepared through the reaction of calcium carbide ($CaC_2$) with hexachlorobenzene ($C_6Cl_6$) or cyanuric chloride ($C_3Cl_3N_3$).[37,53] The reactions were conducted in a planetary ball mill utilizing zirconia jars. These reactions highlight the potential of MSR to facilitate the generation of functionalized carbon structures under mechanochemical conditions.

The mass ratios of 0.9, 2.1 and 5.1 were used to produce carbon nanostructures. The ignition time for the reaction between $CaC_2$ and $C_6Cl_6$ was only 60 s for the mass ratio of 5.1 (***Figure 15***). For other ratios, the MSR ignition times are not reported. The reaction proceeded according to the reaction:

$$3CaC_2 + C_6Cl_6 \rightarrow CaCl_2 + 12C \qquad (7)$$

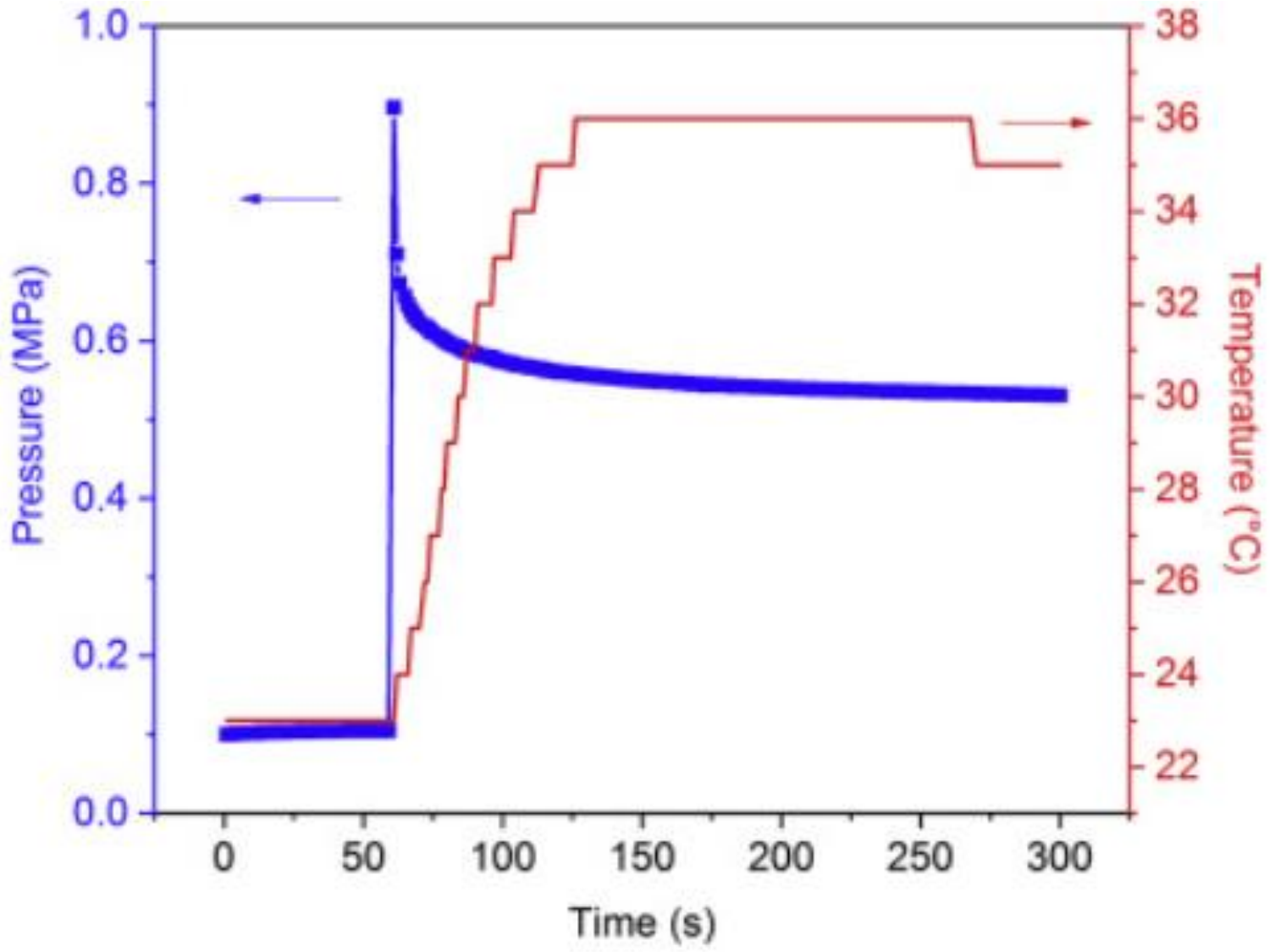

**Figure 15** Pressure and temperature evolution of the mechanically-induced self-sustaining reaction for a mass ratio of $CaC_2/C_6Cl_6$ = 5.1. Reprinted from Ref. [37] Copyright (2018), with permission from Elsevier.

Aa abrupt temperature increase is related to the complete dehalogenation of hexachlorobenzene and an immediate formation of carbon

Another study of mechanochemical synthesis of N-doped carbon nanostructures (abbreviated as "N-Carb" in the study) by using different reagents, namely $CaC_2$ and $C_3Cl_3N_3$, was published in 2019.[53] The ignition time for the reaction between $CaC_2$ and $C_3Cl_3N_3$ ranged from seconds to minutes, depending on the ratio between the reagents following the reaction:

$$3CaC_2 + 2C_3Cl_3N_3 \rightarrow 3CaCl_2 + 2C_6N_3 \quad (8)$$

$CaC_2/C_3Cl_3N_3$ mass ratios of 0.5, 0.7, 2.3, and 4.6 were investigated. The ratio 0.7 is a stoichiometric one. In this case, the reaction took place after 35 min, with a quite high pressure increase. Using the $CaC_2$ over stoichiometry (the ratios 2.3 and 4.6) results in both a much earlier MSR ignition (1 and 5 min, respectively) and also a reduction of the created pressure (**Figure** *16*). The reason why the MSR occurred later when using a larger excess of $CaC_2$ is not reported in the paper; however, it could play the role of an inner additive and thus delay the MSR onset.

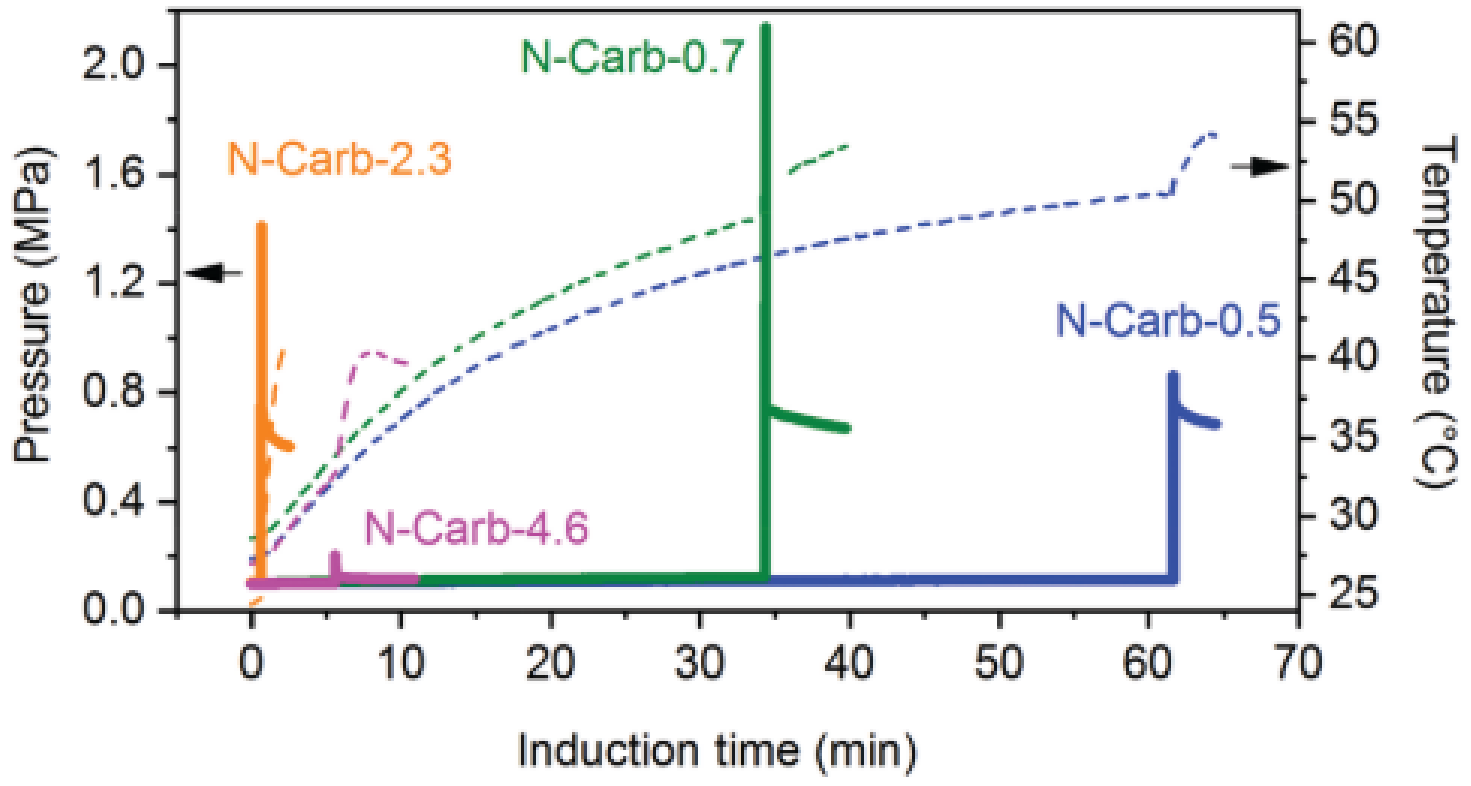

**Figure 16** Changes in gas temperature (dashed lines) and pressure (solid lines) during the mechanochemical synthesis of N-doped carbons. Reactions were followed using an Easy GTM system (Fritsch, Germany). Reprinted with permission from Ref.[53]. 2019. Licensed under CC-BY-NC.

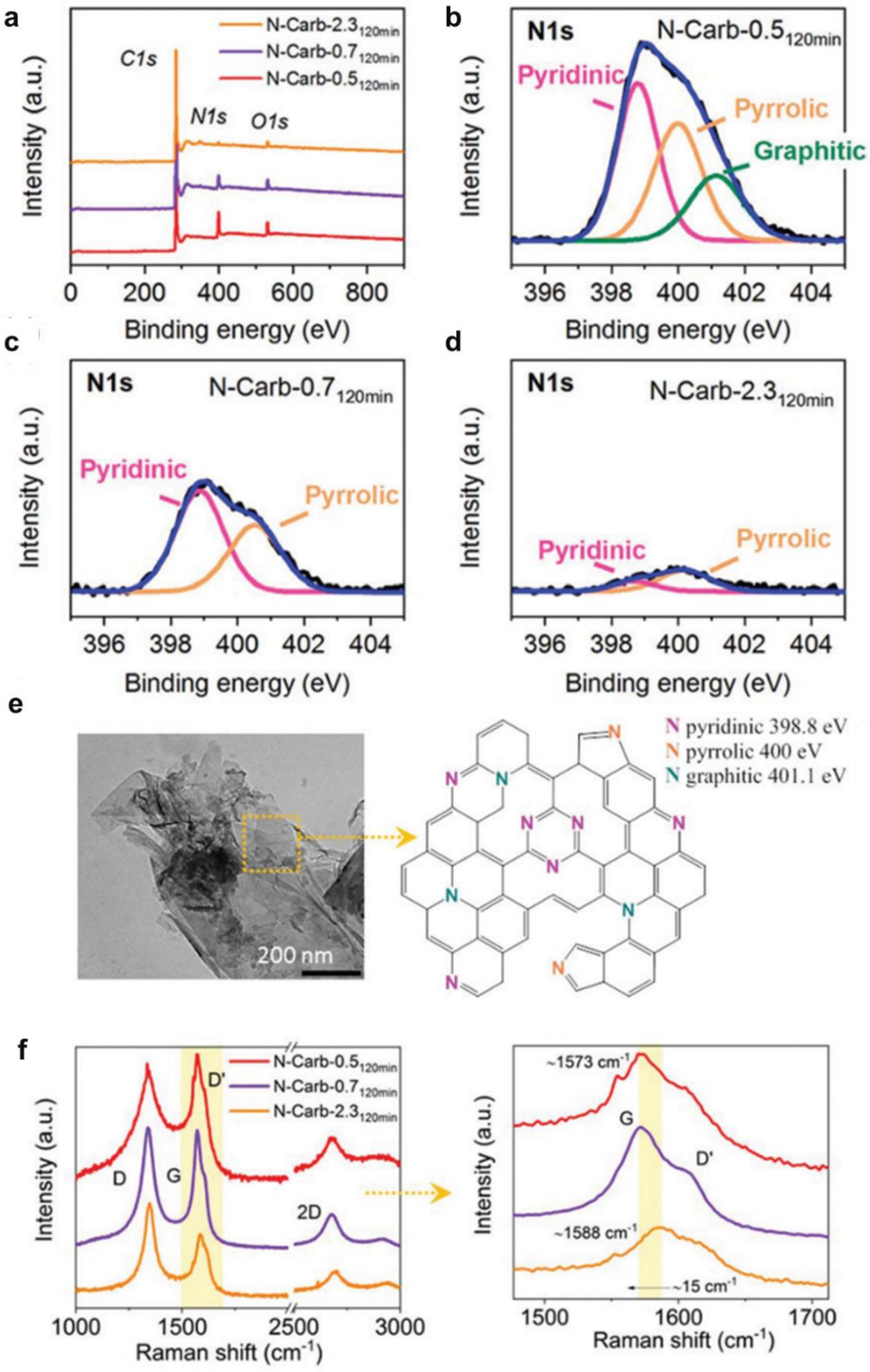

Figure 17 (a) Survey X-ray photoelectron spectra for samples N-Carb- X120 min (X = 0.5, 0.7 and 2.3) obtained in the ball mill. N1s deconvoluted spectra of (b) N-Carb-$0.5_{120\ min}$, (c) N-Carb-$0.7_{120\ min}$ and (d) N-Carb- $2.3_{120\ min}$. (e) Transmission electron microscopy image for sample N-Carb-0.5 and a proposed scheme of a nitrogen-doped graphene layer. (f) Raman spectra for samples N-Carb-$X_{120\ min}$ (X = 0.5, 0.7 and 2.3) obtained in the ball mill and a zoom-in on the G band. Reprinted with permission from Ref.[53]. 2019 licensed under CC-BY-NC.

X-ray photoelectron spectroscopy (XPS) analysis revealed that the intensities of N peaks correlated with the initial mass ratios, namely, the sample with the lowest $CaC_2/C_3Cl_3N_3$ mass ratio exhibits the highest nitrogen content (**Figure *17***a) The sample labelled N-carb-$0.5_{120\ min}$ exhibited three major N 1s components corresponding to pyridinic, pyrrolic and graphitic N. In contrast, only two components, pyridinic and pyrrolic nitrogen were observed in the N-carb-$0.7_{120\ min}$ and N-carb-$2.3_{120\ min}$ (**Figure *17***b-d). Flake-like microstructure was formed as a result of prolonged milling of the powder with the ratio 0.5, as it is shown in **Figure *17***e along with the proposed structure of a nitrogen-doped graphene layer. Raman spectroscopy further confirmed the absence of sp-hybridized carbon, indicating the presence of $sp^2$ hybridization with defects in the aromatic system. These defects are attributed to the substitution of carbon atoms with nitrogen, confirming the doping of N into an aromatic structure (**Figure *17***f).

Apart from porous carbons, also other organic reactions, such as benzil-benzilic acid rearrangement[180] and Buchwald-Hartwig amination[181] were found to run via a MSR pathway. As both these studies utilized the in situ monitoring techniques, they are discussed in detail in the in situ monitoring section (Chapter 4).

## 2.5. Other systems

The systems included in this sub-chapter encompass the ones that do not belong to either of the categories above and for which there are just a few reports on MSRs. As a case study, the one related to metal phosphides was selected.

Although metal phosphides are being used in many applications, such as hydrogen evolution reactions, [182] supercapacitors,[183] or fuel cells,[184] the reports on their preparation by MSR remain scarce. Only one report was mentioned in the review by Takacs,[21] and since then, only one article devoted to this topic has appeared.[52] The transition metals (TM) such as Cr, Co, Mn, Fe, and Ni were mechanochemically processed with phosphorus to prepare $TM_xP_y$. The MSR occurred for the Mn-P and Co-P systems, but the reaction was gradual for the Cr-P, Ni-P, and Fe-P systems. The absence of the MSR pathway for the Cr-P system is peculiar since the adiabatic temperature is high enough to trigger the MSR theoretically. This observation confirms that thermodynamics alone does not govern the MSR ignition.

A specific focus of the mentioned TM-P study was devoted to the role of particle size of the metallic cobalt when preparing cobalt phosphide. Fine Co reacted well through the MSR while using the coarse Co, resulting in a gradual reaction. During the milling of coarse Co and P, intermediate Co-P phases were formed. This suggests that particle size significantly influences the reaction kinetics and the overall MSR process. A similar observation was made for the Cu-S system earlier.[36] Finer particles have a higher surface area to volume ratio, which enhances the contact between reactants and facilitates the initiation and propagation of the reaction. In contrast, coarser particles may require more energy input and time for the reaction to proceed due to their reduced surface area and increased diffusion distances.

Further reactions where MSRs were reported falling into the scope of this review involve the aluminothermic and magnesiothermic reactions to form Ag and Fe nanoparticles and few others. The basic information is summarized in ***Table 7***.

Table 7. Overview of the studies devoted to MSR investigation of other systems since 2002.

| Year | Reagents | Target compounds | Ignition time | Mill type | Ref. |
|---|---|---|---|---|---|
| 2023 | AgCl, Al | Ag | 9 min | Mixer mill | 39 |
| | | | | | |
| 2016 | $FeS_2$, Mg, Al | Fe | 120 min (reduction by Al, in the case of Mg, the reaction was gradual) | Planetary mill | 43 |
| 2005 | Mn, Ni, P, Fe, Cr, P | $Mn_2P$, $Co_2P$, $MnFeP_{0.5}As_{0.5}$ | 55 min ($Mn_2P$)<br>15 min ($Co_2P$)<br>110 min ($MnFeP_{0.5}As_{0.5}$) | Mixer Mill | 52 |

## 3. Factors affecting the MSRs

The occurrence of MSRs is affected not only by the thermodynamic properties but also by milling conditions, and the intrinsic properties of the reagents. In a prior review by Takacs, the relationship between certain milling conditions and the ignition time ($t_{ig}$) was extensively discussed. Key factors such as ball-to-powder ratio (BPR), the inclusion of inert additives, stoichiometric variations, the effect of milling interruptions, and milling frequency were analyzed in detail.[21] Building upon this foundation, the present review revisits some of these earlier findings while integrating additional factors that have emerged as critical influencers of MSR behavior. These include the impact of reagent morphology, the selection and configuration of milling devices, and the role of process control agents (PCAs) in modulating reaction dynamics. The influence of some factors such as the effect of free volume in the jar, breaks duration or the exact influence of flaky morphology was not studied at all so far. A comprehensive figure showing the factors influencing the MSR ignition is provided below (***Figure 18***). The data briefly summarized in this chapter provide a comprehensive and updated perspective on the mechanistic intricacies of MSRs.

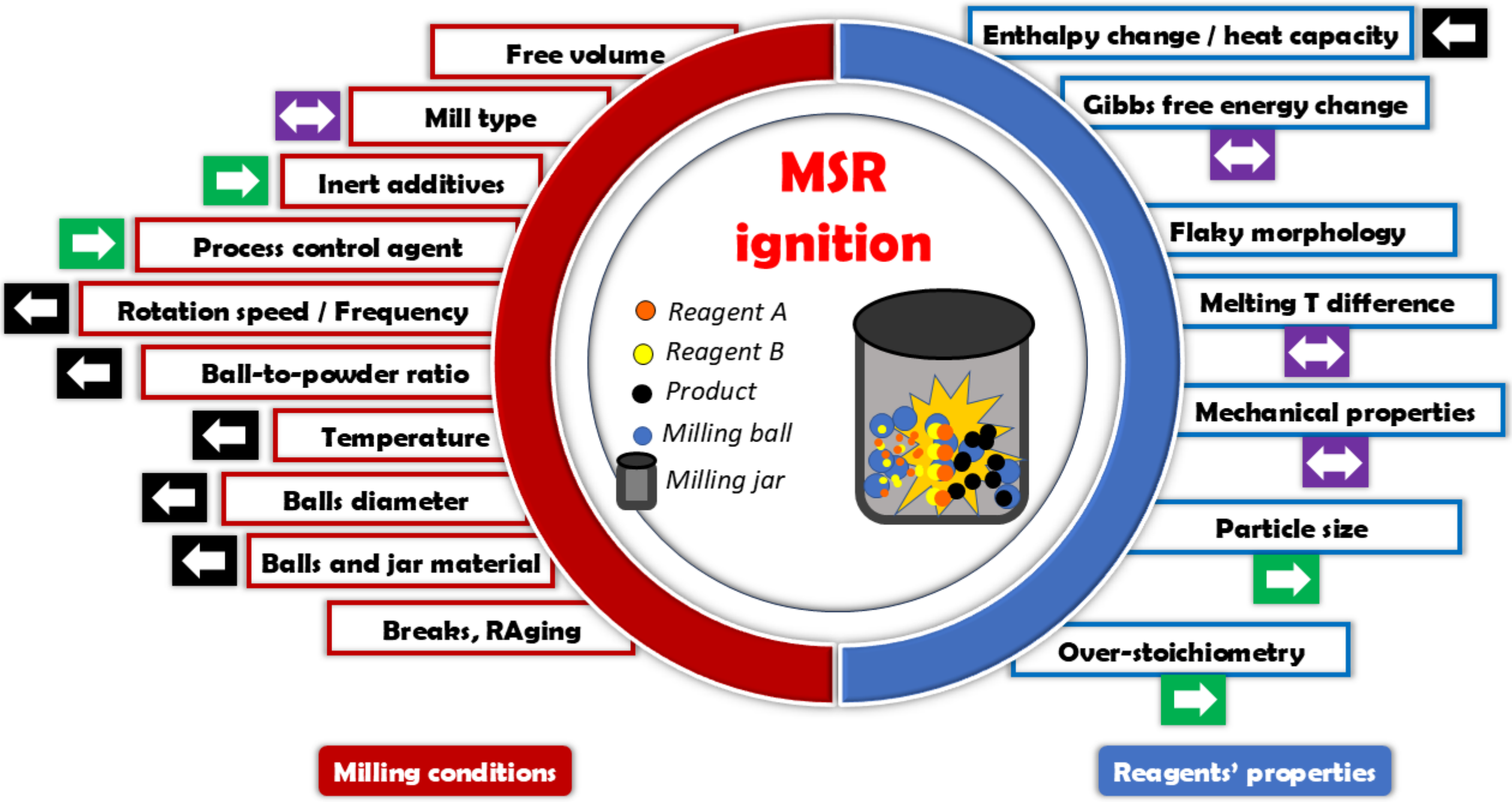

Figure 18 Factors affecting the ignition time of mechanically induced self-propagating reactions. If the general influence of the given factor is clear from relevant literature, the relationship with MSR ignition time is provided therein (*green arrow- direct*, i.e. increasing this factor's value leads to a delay of MSR ignition, *black arrow- inverse*, i.e. increasing this factor's value leads to an earlier MSR ignition, *purple arrows- not clear* but studied in literature, *no arrow- not studied*).

The factors influencing the MSR can be categorized into **milling conditions** and **reagents' properties**, with each playing a pivotal role in determining the $t_{ig}$ and reaction pathways.

**Milling conditions**, such as frequency, temperature, ball-to-powder ratio (BPR), spinning rate, inert additives, temperature and processing control agent (PCA) influence the energy input, collision dynamics and reaction kinetics. The following factors exhibit an inverse relationship with $t_{ig}$, i.e. the larger the value of the factor, the sooner the MSR occurs:

a. rotation speed and frequency; higher operational frequencies and spinning rates increase collision intensity, effectively lowering $t_{ig}$, as seen in Ti(BCN),[185] TiC,[63] $Cu_2S$,[164] or ZnSe.[35]
b. ball-to-powder ratio (BPR): higher BPR generally enhances collision frequency and energy transfer, reducing $t_{ig}$ as demonstrated in systems like ZnSe,[35] $Fe_{4.5}Ni_{4.5}S_8$,[44], and NiAl intermetallic.[54] In Ti-B, however, the MSR were observed after 73, 34, and 40 minutes for BPRs of 10:1, 15:1, and 20:1, respectively.[109] The optimal BPR was determined to be 15:1. This anomaly may be attributed to the free volume within the milling jar, which plays a significant role in facilitating effective collisions. It is well established that the highest collision efficiency is achieved when the vial is filled to two-thirds of its total volume.[186]
c. temperature; elevated temperatures decrease $t_{ig}$ by accelerating diffusion and reaction kinetics as show for ZnSe[187], the MSR ignition time almost reaching zero when T was increased to melting temperature of zinc- this was observed for ZnS.[187]
d. ball diameter; larger balls cause MSR to occur sooner and if the balls are smaller than 5 mm, the reaction becomes gradual. This was reported for the Ti-C system.[188]

e. balls and jar material; using denser milling media speeds up the MSR ignition. [35]

A direct relationship i.e., delaying the MSR ignition time or turning the reaction into a gradual one was observed with the following factors:

a. utilizing process control agent (PCA); the presence of PCA e.g, in WC-MgO,[91] and Ti-BN[189] systems, prolonged $t_{ig}$ by reducing cold welding and stabilizing the milling environment; if the content of PCA is too high, the reaction changes into a gradual one.
b. inert additives; their presence, e.g, in the system $TiO_2$-$B_2O_3$-Mg-C [190] suppresses MSR by diluting reactants, thereby extending $t_{ig}$ or changing the reaction into a gradual one.

To assess the effect of different mill types , an attrition and a planetary ball mill were used to monitor the MSR between Ni and Al metals to form intermetallic alloys.[191] The authors claim that MSR occurred only in the attrition mill, whereas in the planetary mill, the reaction was gradual. It is true that no abrupt pressure increase was evidenced during planetary milling, however, sometimes the sensor fails to properly catch it. The XRD patterns are quite similar, showing a sudden appearance of the intermetallic phase also in the planetary mill.

**Reagent's properties,** including morphology, crystal size, mechanical properties, thermodynamic properties, and stoichiometry also affect the occurrence of MSR. There is a linear relationship with $t_{ig}$ and particle size, e.g in the system of NiAl,[54] Co-P,[52] and ZnS.[187] The effect of pre-milling in the system of CdSe and ZnSe [159] contributes in a similar manner, as it reduces the particle size of the reactants and results in earlier MSR ignition.

In literature, there is a theory regarding the influence of reagents, melting temperature difference and ignition time- the larger the difference, the later the MSR occurs.[47] This theory was tested only on a very small group of reaction systems for now, and its validity needs to be tested on a more general level.

In relation to mechanical properties, a theory regarding the hardness of metals and chalcogen was postulated in the pioneering work on metal chalcogenides preparation via MSR, namely that the reaction occurs most quickly if the metal is soft and the chalcogen is hard.[24] A clear relationship between Gibbs' free energy change and MSR ignition time for multiple metal chalcogenides could not be identified.[45]

The stoichiometry of the reagents plays a decisive role in the MSR process. The MSR usually occurs in a certain range of reagents' concentrations. If this is not respected, the reaction pathway changes into a gradual one, as the excessive reagent plays the role of an inert additive. This can be exemplified in the Al-$B_2O_3$-C system.[192] The aluminothermic reaction proceeded gradually at lower Al content (up to 2 mol). However, increasing the Al content to 3 mol changed the reaction to a MSR. When the Al content was further increased to 3.3–4 mol, the aluminothermic reaction generated sufficient heat to activate the endothermic carbothermic reduction, allowing the following two reactions to occur simultaneously:

$$2B_2O_3 + 11/3Al + 3/2C = 11/6Al_2O_3 + B_4C + 1/2C \quad (9)$$

$$2B_2O_3 + 4Al + C = Al_2O_3 + B_4C \quad (10)$$

At an Al content of 7 mol, the aluminothermic reaction became gradual again, with excess Al acting as an inert additive. A large amount of additive generally reduces $T_{ad}$ below the threshold required for MSR initiation.[21]. A similar behavior was observed in the $WO_3$-Al-C system,[193] and in the reaction between $MoSi_2$ and W.[121] The composition of the Ti-C system and the input energy significantly influence the MSR. The MSR occurs when the composition falls within the range of $Ti_{45}C_{55}$ and $Ti_{72}C_{28}$, with the collision energy higher than 0.027 J and lower than 0.155 J. Outside this range, when the composition lies between $Ti_{20}C_{80}$ and $Ti_{45}C_{55}$ or between $Ti_{72}C_{28}$ and $Ti_{80}C_{20}$, a gradual pathway is observed. This is attributed to the insufficient or excessive collision energy.[75]

*Table 8. The factors affecting the MSR and* ***their relationship*** with $t_{ig}$.summarizes the studies devoted to investigating different factors affecting the MSR published since 2002, together with the identified relationship with ignition time, $t_{ig}$ (if available).

**Table 8.** The factors affecting the MSR and their relationship with $t_{ig}$.

| Year | Studied system | Mill type | Studied factor | Relationship with $t_{ig}$ | References |
|---|---|---|---|---|---|
| 2025 | Buchwald-Hartwig amination | Resonant acoustic mixer | Jar filling ratio | Inverse | 181 |
| | | | Acceleration | Inverse | |
| 2024 | $Ti_3AlC_2$ | Planetary mill | Stoichiometry | Inverse until 1.2 % Al, then direct | 76 |
| 2024 | Metal-chalcogen | Planetary mill | Hardness | No correlation | 45 |
| 2022 | Ni-$Cr_3C_2$ | Not mentioned | Stoichiometry (excess Mg) | Inverse | 77 |
| 2021 | Zn-S | Mixer mill | Particle size | Direct | 187 |
| | | | Temperature | Inverse | |
| 2020 | Fe-Ni-S | Planetary mill | BPR | Inverse | 44 |
| 2019 | Ti(BCN) | Planetary mill | Milling speed/frequency | Direct | 185 |
| 2018 | $CaWO_4$-Mg | Planetary mill | Stoichiometry | Certain range affect $t_{ig}$ | 194 |
| 2017 | Cu-S | Planetary mill | Copper morphology | Direct | 164 |
| 2015 | Cd-Se, Zn-Se | Planetary mill | Particle size | Direct | 159 |
| 2014 | Ti-C | Mixer mill | Stoichiometry | Certain range affect $t_{ig}$ | 75 |
| 2014 | $TiB_2$ | Planetary mill | BPR | 10:1, 15:1 and 20:1, BPR 15:1 is optimum | 109 |
| 2014 | $TiO_2$-$B_2O_3$-Mg-C | Planetary mill | Inert additives | Direct | 190 |
| 2014 | Sn-Zn-Se | Mixer mill | Composition $(1-x)(Sn+Se)+x(Zn+Se)$ | Direct (as per Zn content) until 75 at. %, then inverse | 195 |
| 2013 | ZnSe | Planetary mill | Spinning rate | Spinning rate cubed is proportional to the inverse of $t_{ig}$ | 35 |
| | | | BPR | Inverse | |
| 2013 | $MoSi_2$ | Mixer mill | Stoichiometry | Certain range affect $t_{ig}$ | [122] |
| 2013 | Ti-C | Planetary mill | Milling speed | inverse | [188] |
| | | | Ball diameter | inverse | |
| 2012 | Ni-Al | Planetary mill<br>Attritor mill | Milling devices | MSR in attritor, gradual in planetary mill | [191] |
| 2011 | Ni-Al | Mixer mill | BPR | Inverse | [54] |
| 2011 | Ti-Si/Ge | Planetary mill | Stoichiometry | Certain range affect $t_{ig}$ | [134] |
| 2010 | $MoSi_2$ | Vibratory mill | Additives (W and Al) | W: linear, Al: inverse x=0-0.1 mol, linear, x>0.2 | [121] |
| 2010 | WC-MgO | Planetary mill | PCA | Direct | [196] |
| 2010 | WC-MgO | Planetary mill | | | |
| 2009 | $WO_3$-Al-C | Planetary ball mill | Stoichiometry | Inverse | [193] |
| 2007 | Ti-C | Magnetically controlled mill | Milling speed | Inverse | [63] |
| | | | Inert additives | Linear | |
| 2005 | Ti-C | Magnetically controlled mill | Properties | No specific correlation with $t_{ig}$; explain more the reaction pathway, either MSR or gradual | [74] |
| 2005 | Ti-C | Magnetically controlled mill | Stoichiometry | Certain range affect $t_{ig}$ | [64] |
| 2005 | Co, P | Mixer mill | Particle size | Direct | [52] |
| 2004 | Al-$B_2O_3$-C | Planetary ball mill | Inert additives | Direct | [192] |
| 2004 | $Fe_2O_3$-Al | Vibratory mill | Frequency | Inverse | [197] |
| 2003 | Ti-BN | Planetary mill | PCA | Direct | [189] |
| 2003 | Ti-C | Vibratory mill | Inert additives | Direct | [198] |

## 4. *In situ* monitoring of the MSR

The combustive character of MSR, driven by the exothermicity of the reactants, has fascinated researchers for a long time, and they have tried to develop different methods to be able to track the process as closely as possible. This was initially done by implementing thermocouples,[69,135,177,199–201] later on by using temperature and pressure sensors capable of monitoring what happens inside the jar during milling,[44,45,94,113] and recently, by applying time-resolved *in situ* monitoring methods[202] (TRIS) as exemplified e.g., in.[38,39] Thermocouples and sensors are advantageous tools for probing MSR behavior, particularly for determining ignition time $t_{ig}$, which is often the primary parameter of interest, a measure of reactivity. Spontaneous ignition of the MSR process is associated with specific stages of microstructural refinement within the reactant phases and increasing structural disorder. However, the fundamental mechanisms driving reaction ignition, propagation mode, and propagation rate remain incompletely understood, warranting further study to optimize the control and predictability of these processes.

In many studies, the MSR occurrence is reported based just on the fact that there is a sudden change in the *ex situ* XRD patterns. While that can be the first signal, it does not mean that the reaction actually follows the MSR pathway. Definite proof of its occurrence can be obtained by *in situ* temperature and/or pressure monitoring. This was shown in the past e.g. during the preparation of metal sulfides and selenides [45,164], Ti-C [203], or the production of porous carbon and N-doped porous carbon [37,53]. Some researchers developed their own set-up mainly for temperature monitoring, but there are also commercially available solutions from companies that provide the milling equipment, such as EasyGTM [204]or GrindControl.[205] The commercial solutions have more limitations, such as low collection frequency and high cost, and in the case of EasyGTM, the sensor is not transferable. It can only be purchased and used together with the specifically engineered milling jar. These drawbacks hamper their proper utilization for precise detection of MSR processes. Moreover, the MSR occurrence cannot be observed on the temperature curve due to the fact that the thermal sensor is placed quite far from the inside of the milling chamber. Thus, the MSR ignition can be tracked using only a more expensive pressure sensor in the commercial solutions. When the MSR occurs, heat generation and gas expansion lead to sudden pressure changes. Pressure sensors capture this almost instantaneously, providing real-time insight into ignition and propagation. With the sensitivity to subtle pressure changes, pressure sensors detect even minor fluctuations in pressure that might precede or coincide with ignition, revealing early signs of a reaction starting or intensifying. To make this possible also for temperature monitoring, our group has developed a transferrable temperature sensor for a planetary ball mill, which collects the data every 80 ms, featuring a resistance tolerance of ±1% at 25 °C and a beta value tolerance of ±1%, ensuring high-precision temperature measurements.[201] This invention allows for the precise detection of MSR by *in situ* temperature monitoring, identifying a temperature spike at the moment of MSR. The sensor is in direct contact with the milling environment. The patent application regarding this invention was filed in Slovakia in 2023.[206]

An investigation of the MSR using the pressure and temperature sensors is useful for providing information about the ignition time. However, it does not provide the spectroscopic proof. The MSRs are also accompanied by a flash of light inside the milling chamber, which can be seen when a transparent one is used. This was accomplished, e.g. in the study of TaC and TiSi synthesis in a horizontal oscillation mill using infrared thermometry and a transparent fused quartz chamber.[199,207] The occurrence of the MSR and the accompanying temperature change in the TaC system is presented in ***Figure 19***. Ignition is thought to occur on the vial's base due to either intense attrition or the impact of milling balls. While this approach does not provide quantitative phase or microstructural analyses, it facilitates qualitative observation, such as the appearance of bright spot particles possessing elevated temperatures (***Figure 19***a -e) and the further onset of the reaction in neighboring particles and propagating all across the whole volume of the sample.

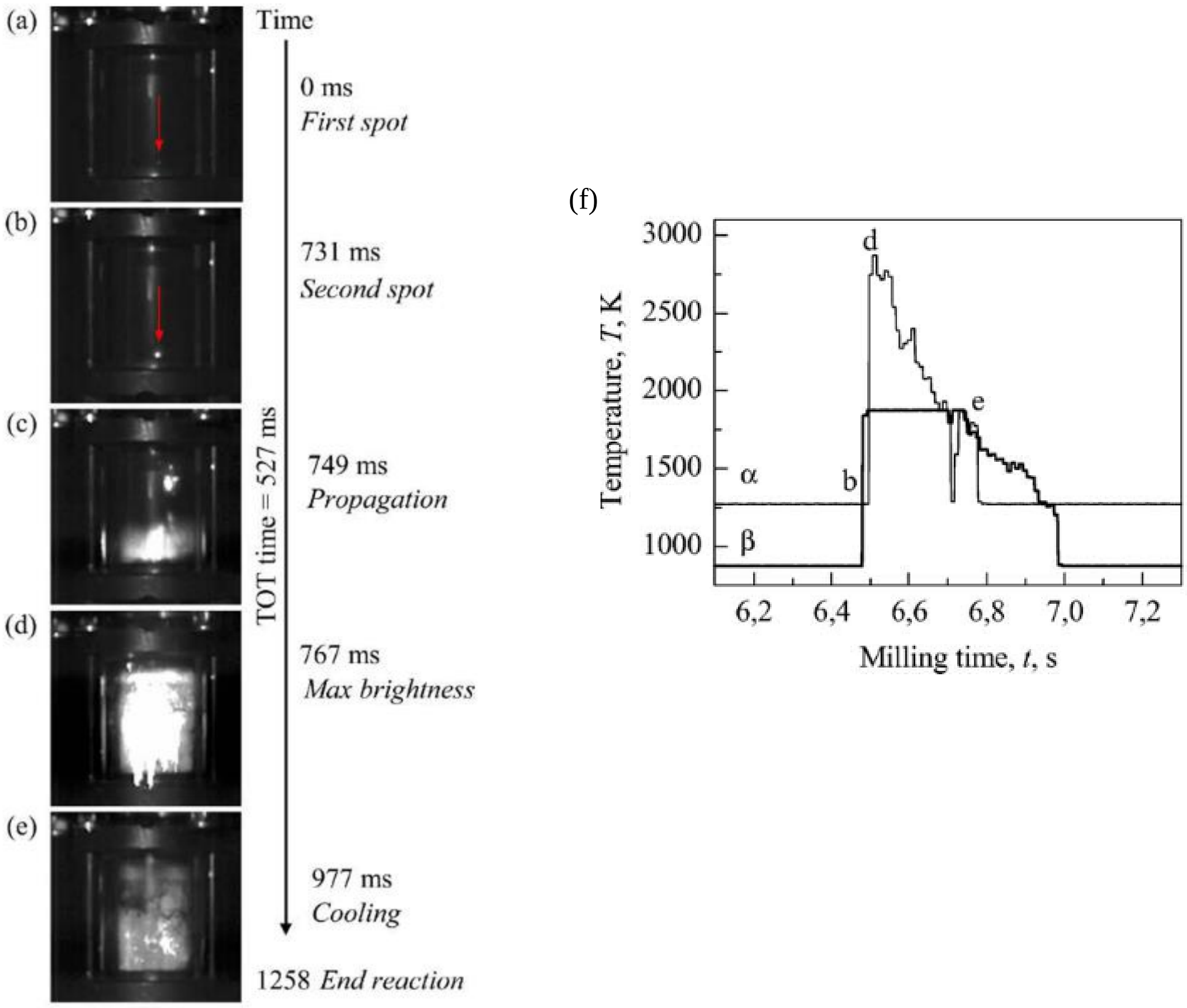

**Figure 19** High-speed image sequence of the reaction of the Ta:C=50:50 mixture in the horizontal oscillation mill coupled with the related temperature recording obtained by the infrared thermometers. Reprinted with permission from [199], 2004, Springer Nature.

The technical development of TRIS monitoring techniques providing information directly during milling paved the way for an in-depth mechanistic understanding of mechanochemical reactions.[208] The most commonly used TRIS methods with a high information content are Raman spectroscopy and X-ray diffraction. However, TRIS characterization techniques have been only scarcely utilized for the investigation of MSRs until now.

The reason why MSR monitoring via TRIS techniques remains on the periphery of research until now lies in the fact that it usually takes a large amount of energy to ignite the MSR process, and thus, usually, it is conducted in planetary ball mills where chambers are made of dense metallic or ceramic materials. However, to be able to spectrally track the MSR, the reaction has to be conducted in a transparent milling jar, most often prepared from the polymer. To the best of our knowledge, there are just two studies that employed the TRIS methods to investigate the MSR processes, and they are briefly reviewed below.

In the first one, the synthesis of silver nanoparticles from silver chloride by using aluminum as a reductant [39] was effectively tracked using *in situ* X-ray diffraction monitoring with synchrotron radiation and thermography. The XRD patterns were collected every second during exposure. Milling was conducted in a shaker ball mill with a 3D-printed polycarbonate jar and four agate balls. The results point to the reaction releasing considerable heat after a certain amount of time, but the formation of the product is not immediate, thus not aligning with the general picture of MSR. The authors monitored the changes in relative amounts of AgCl, Al (reagents), and Ag (product), which were determined by the normalization of the scale factors from Rietveld refinements for each phase (***Figure 20***a). Until 9 minutes, the relative amount of both AgCl and Ag increased (a small amount of Ag was already formed until that time), while that of Al decreased. In the case of AgCl, this increase is most probably connected with the growth of AgCl crystallites prior to the major phase of the reaction. After 9 minutes, the temperature started to increase, peaking at around 10 minutes and returning to normal after around 11 minutes, as corroborated by the thermography data. As for relative amounts, that of silver significantly increased within these two minutes and continued to increase also after the termination of the temperature event. The one of AgCl decreased only during the temperature event and started to increase again afterwards. One of the Al was not influenced during heat formation. The conversion was not complete, as AgCl can still be traced at the end of the process. The XRD patterns in ***Figure 20***b show, it seems that there is a rapid formation of Ag after 9 min, but the reaction takes longer and is not instantaneous. When the temperature increase occurs, the reaction speeds up and is significantly advanced within the following 2 minutes, but does not reach the state of classical MSRs (almost quantitative completion). Nevertheless, this study shows a great way of investigating MSR fundamentals, which can be applied in future studies.

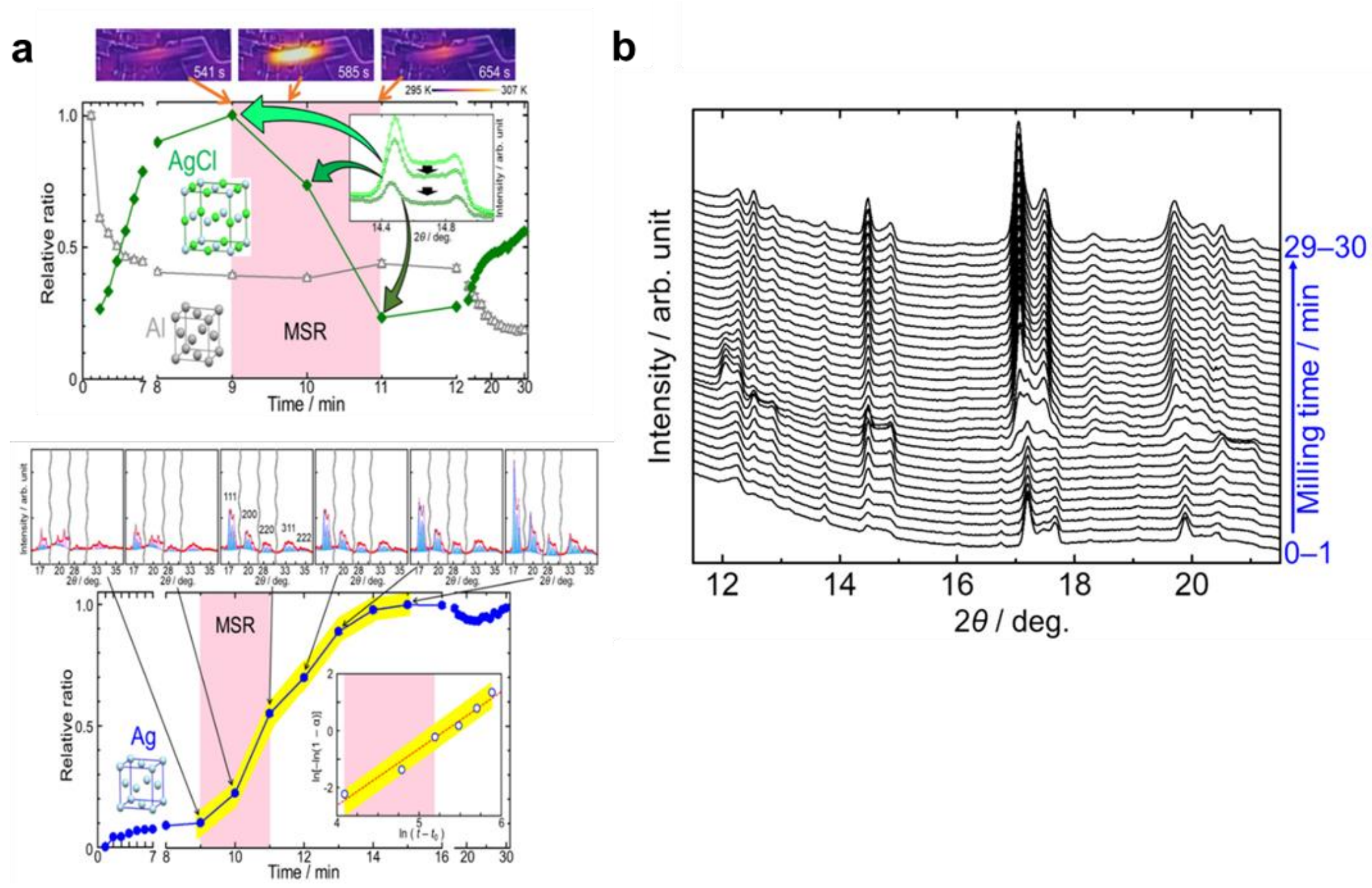

Figure 20 (a) Time evolution of the relative amounts of reactants (AgCl and Al, top) and product (Ag, bottom), (b) *in situ* XRD monitoring results. Reproduced from Ref. [39] with permission from the Royal Society of Chemistry.

In a comprehensive investigation of the ZnS system by the group of Weidenthaler [38] the *in situ* XRD monitoring was utilized to examine the milling process of elemental Zn and S powders in a shaker mill. Initially, it was necessary to develop a specific milling chamber made of steel, with a transparent window made of polymer, so that the X-rays could penetrate inside and also to modify the construction of the mixer mill. This was achieved by the same research group earlier.[209] The study on Zn-S revealed the formation of the metastable hexagonal ZnS phase at the moment of MSR, transitioning to the cubic phase almost immediately. This is demonstrated by showing a section of time-resolved XRD patterns in the 2 theta range 3.5–4.1°.  The pressure detection was influenced by the milling frequency and the jars. Two types of jar holders were used in this experiment, namely the commercial and modified one. In both cases, the MSR ignition time decreased with the increase of the milling frequency, however, whereas 23 Hz was a sufficient frequency to ignite the MSR in the commercial jar holder, no reaction as observed when using the modified one. Similarly, the MSR ignition times were always delayed at higher frequencies when using the modified jar holder in comparison with the commercial one. Thus, controlling the impact energy and efficiency of energy transfer to the reactants via changing milling frequency plays a crucial role in the formation and stability of ZnS phases.

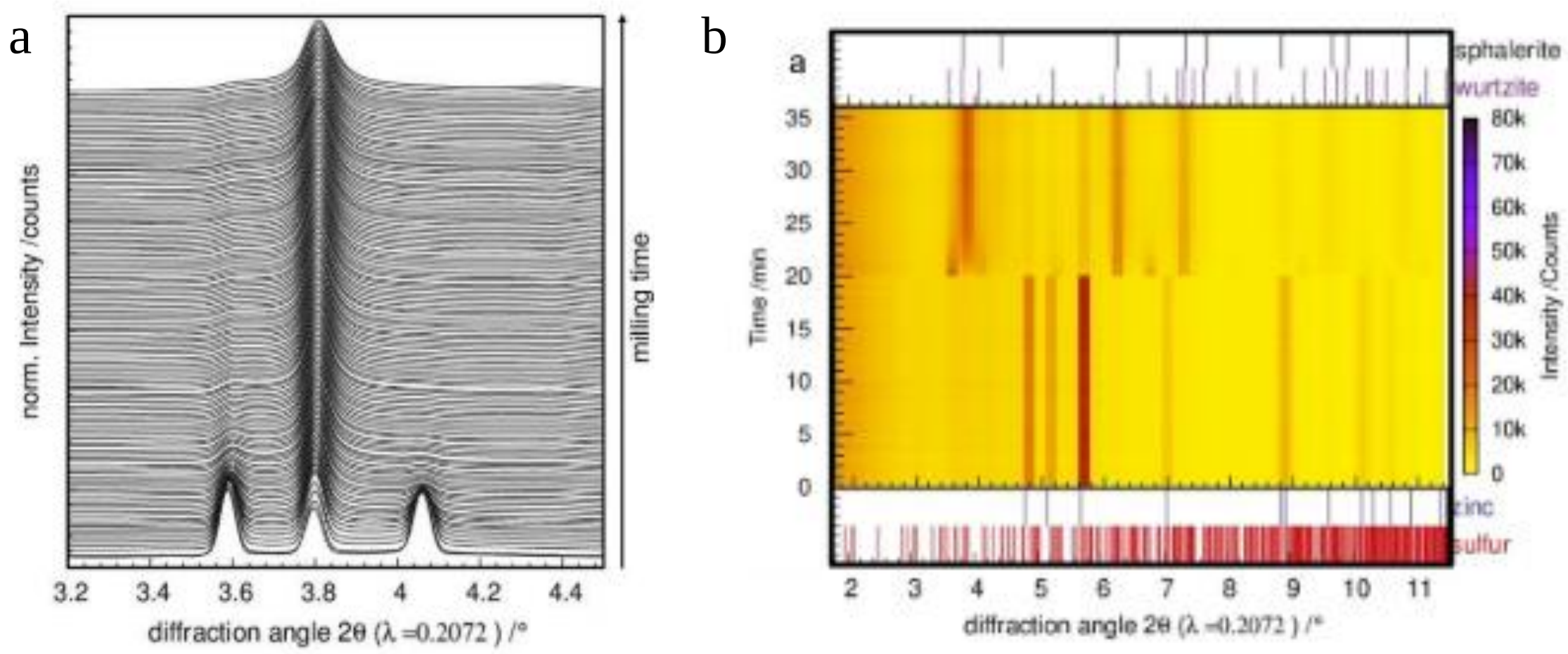

**Figure 21** (a) Time-resolved evolution of the main reflections of the hexagonal ZnS (wurtzite) obtained immediately after the MSR occurrence into the cubic sphalerite upon milling at 30 Hz , (b) *In situ* XRD pattern of a Zn:S 1:1 reaction mixture milled with 30 Hz. Adapted from Ref. [38] licensed under CC-BY.

The TRIS methods were also used for the investigation of MSR in the field of organic chemistry. Namely, benzilic acid rearrangement (***Figure 22***) was investigated using *in situ* X-ray diffraction, Raman spectroscopy and temperature sensing.[180] The reaction was carried out  in a mixer mill using a PMMA jar and two tungsten carbide balls. The amount of introduced KOH governed the occurrence of MSR, as shown in ***Figure 23***.

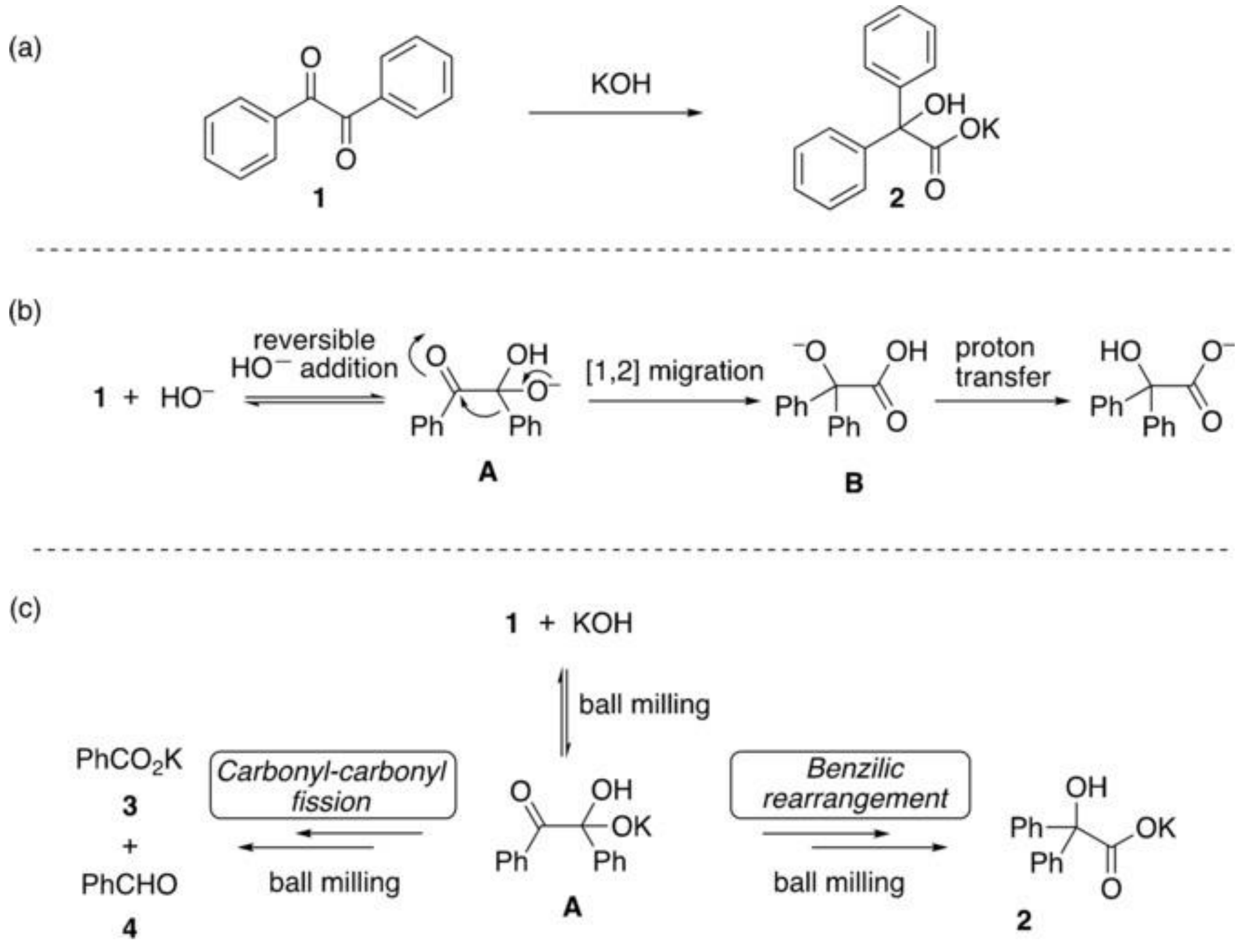

**Figure 22** a) Benzil–benzilic acid rearrangement in the presence of potassium hydroxide. b) The classically accepted mechanism for the benzil–benzilic acid rearrangement in the presence of hydroxide ions in solution. c) Mechanochemical base-induced benzilic rearrangement and base-induced carbonyl–carbonyl fission to generate products **2**–**4**. Reprinted from Ref. [180] licensed under CC-BY.

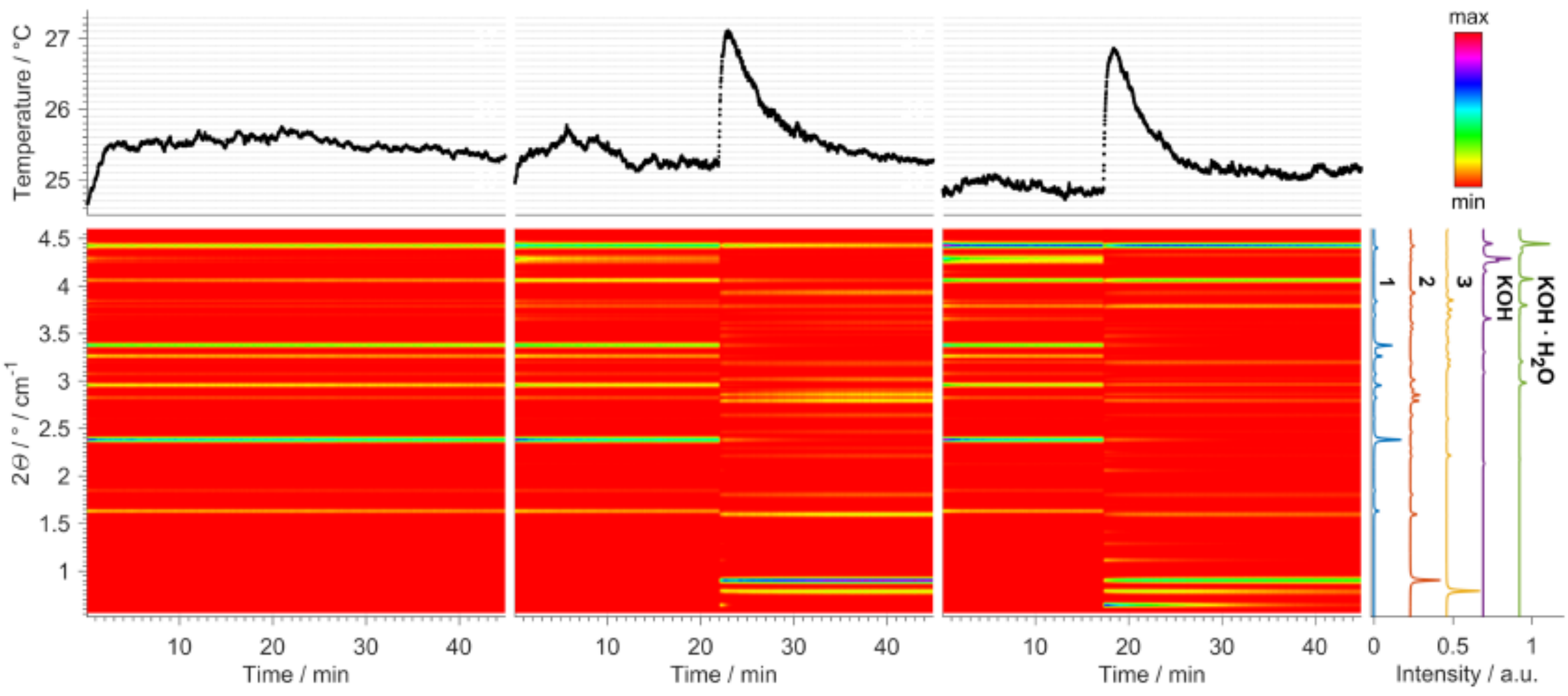

**Figure 23** Time-resolved X-ray diffractograms and temperature profiles for reactions of benzil (**1**) and (left) KOH (1.0 equiv); (middle) KOH (2.0 equiv) and (right) KOH (3.0 equiv) milled at 30 Hz. PXRD patterns of benzil (**1**), potassium benzilate (**2**) and potassium benzoate (**3**) KOH and KOH・$H_2O$ are given on the right side of the 2D plots. Adapted from Ref. [180] licensed under CC-BY.

As shown in ***Figure 23***, with the increase in the amount of KOH (in equivalent), the duration of the activation period was reduced. When using 1 equivalent, no MSR was evidenced, while when using 2 and 3 equivalents, the MSR ignition time was 22 and 17 min, respectively. The MSR occurrence, similarly to inorganic systems, is evidenced by the abrupt change in the XRD pattern. The presence of potassium benzilate (**2**) confirms the successful molecular arrangement. Benzoate, as the minor product, exists due to a competitive carbonyl carbon–carbonyl carbon bond fission in benzil. Moreover, a short-lived intermediate at 2[illegible]° was observed in excess KOH (2.0 and 3.0 equiv.).

Another very recent example of an *in situ* monitoring of a MSR-like event in an organic reaction is a Buchwald-Hartwig amination[181] performed using resonant acoustic mixing (RAM), a scalable and ball-free alternative to a classical mechanochemical equipment.[210] The authors investigated mainly the effect of filling ratio, RAM acceleration and amount of liquid additive. The scheme of the reaction is in ***Figure 23*** and the *in situ* Raman monitoring data with the corresponding temperature profiles of the reactions performed under different filling ratios are given in ***Figure 23***.[181]

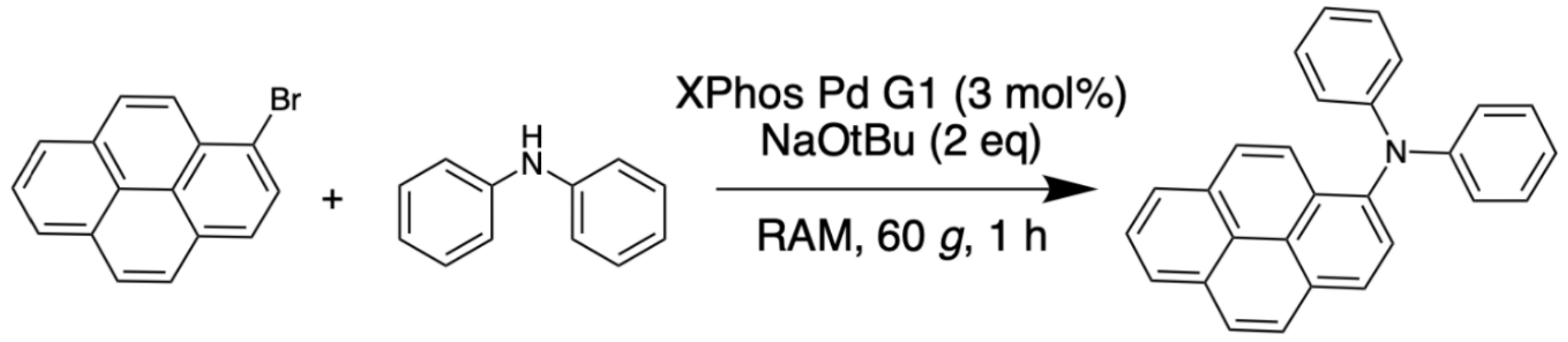

**Figure 24** Scheme of a Buchwald-Hartwig amination of 1- bromopyrene and diphenylamine with the experimental conditions.[181]

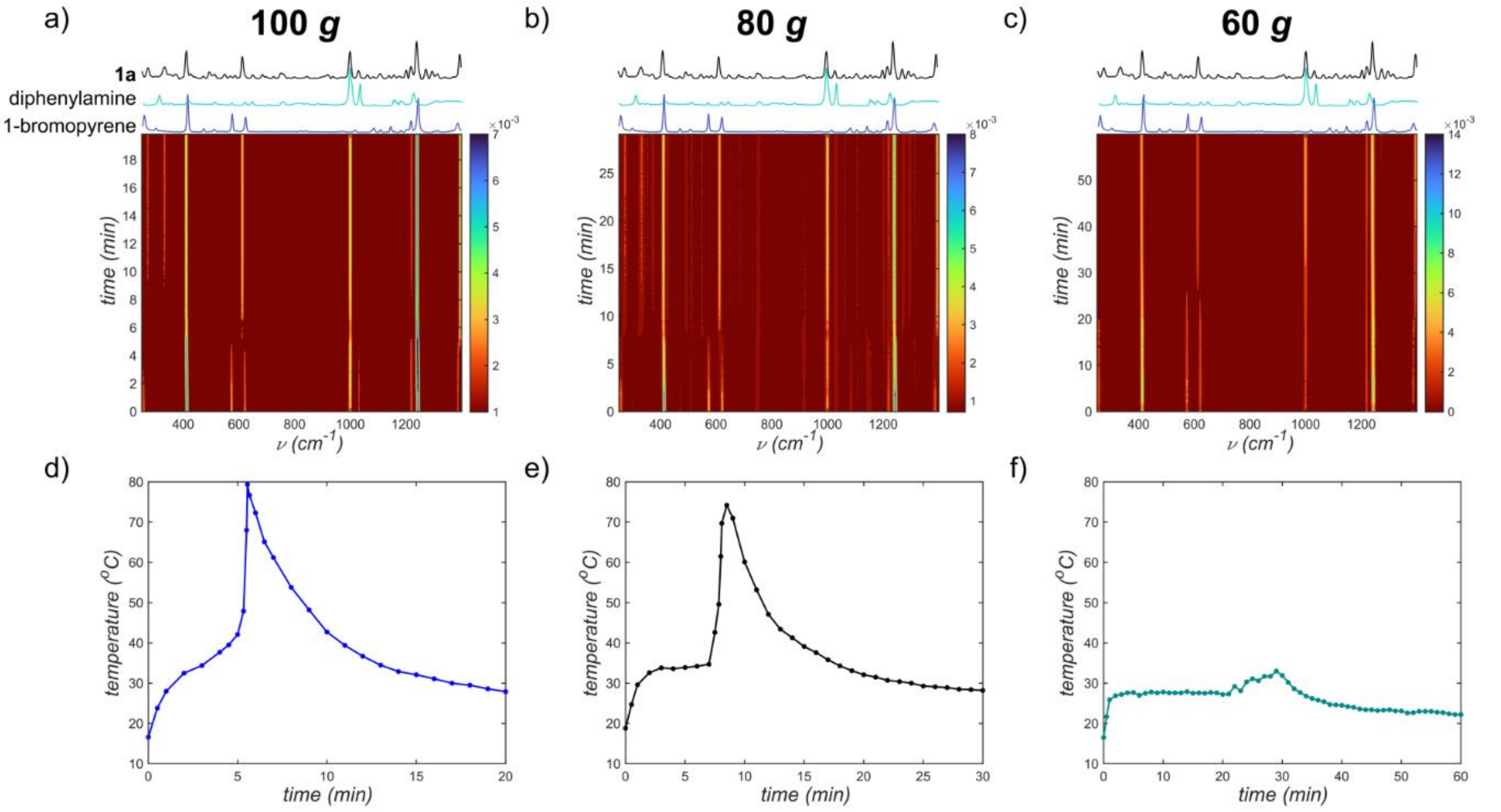

**Figure 25** Results of real-time thermal and Raman spectroscopy monitoring of the RAM Buchwald-Hartwig amination of 1-bromopyrene and diphenylamine: (a-c) Raman spectroscopy time-resolved water-fall plot for the reaction run at a) 100 g, b) 80 g , and c) 60 g accelerations; (d-f) in situ thermography data for reactions run at d) 100 g, e) 80 g, and f) 60 g. Reaction conditions: diphenylamine (4.5 mmol), 1-bromopyrene (4.5 mmol), NaOtBu (9 mmol), XPhos Pd G1 (3 mol%), THF (0.25 µL/mg), jar filling ratio 80 %.[181]

Both *in situ* Raman monitoring data and temperature curves clearly show the occurrence of an abrupt temperature increase after a certain time, however, this process seems to be longer than usually evidenced in the case of classical MSRs. On the other hand, the exact detection of MSR is determined by the set-up used for monitoring. In this case, the placement of the temperature sensor might cause a delay and thus, the sub-second MSR course is reflected later in a slower manner due to a limited heat transfer. Nevertheless, the reaction occurrence is governed by the RAM acceleration, as the higher g results in a sooner reaction onset. The Raman bands are changed from those of reagents into products very rapidly. Apart from acceleration, also jar filling ratio was found to be very important, since the temperature event occurred after around 4 minutes when using 80% filling ratio, and was postponed to around 35 min with a much smaller temperature increase when conducting the reaction at 40% filling ratio.

## 5. Conclusions and future perspectives

The recent revelations in researching MSRs within the realm of mechanochemistry underscore the unique feature of this process to yield the products in a more energy-efficient process compared to the gradual reactions. This is because the MSR takes place significantly earlier than the completion of the gradual process. Following the mechanical activation period, the MSR is usually ignited and completed in a sub second range. More proofs of that have been recently evidenced mainly by utilizing time-resolved *in situ* monitoring techniques and introducing the sub second range temperature monitoring sensor. The former confirm the spectroscopic evidence for that, while the latter clearly shows the drawbacks of commercially available tools to precisely detect the super-fast MSR process by temperature monitoring. The fact that MSR is accompanied by a flash of light has now been confirmed multiple times. The MSRs were confirmed to occur both under air and inert atmosphere. The scope of the reactions susceptible to MSR has been significantly broadened within the last twenty years, including the organic transformations. Controlling the course of MSRs is so far possible only to a limited extent via modifying either milling conditions or reagents' properties. The former approach is more common, and some correlations are already well-known. The latter is much more challenging due to the complexity and large number of factors that can possibly influence the MSR from the reagents' properties perspective and the scarcity of the available reports.

There is a definite need to understand MSRs on more general and fundamental level, specifically to identify the rules that govern their occurrence and course. This is specifically relevant for the properties of the reagents, as there are strong indications that the fine particle size favors the MSR occurrence. But what about the exact limits of particle size below which the MSR occurs and above which it does not occur? Is it specific for each reaction under study? What about the influence of morphology and crystal structure? These are very basic questions just related to a small portion of factors that influence MSRs. Answering them should be the focus of the forthcoming research to understand MSRs more fundamentally and solve more general questions in the future such as whether it will be possible to use MSRs as an established reaction pathway to reach advanced materials and what kind of precautions should be taken to apply them on a larger scale. With relation to mechanochemistry field, the ignition of MSRs via utilizing resonant acoustic mixing[210] twin screw extrusion,[211] or even the photo-, thermo-, electro- and sono-mechanochemistry,[212] i.e. techniques in the current focus of the mechanochemical research, under ball-free conditions, should be investigated.

Mechanochemistry is a green and environmentally sustainable approach already established within chemical and materials science research, as often corroborated by great results in green metrics calculations. However, further improvements are still being sought, and utilizing and understanding MSR on a more general level will surely bring further enlightenment within this fascinating discipline.

## Declaration of generative AI and AI-assisted technologies in the manuscript preparation process.

During the preparation of this work the authors used ChatGPT version 5.2 to prepare a graphical abstract of the manuscript. After using this tool, the authors reviewed and edited the content as needed and take full responsibility for the final image.

## Acknowledgement

This study was supported by the Slovak Research and Development Agency under Contract No. APVV-24-0353. The support of Grant Agency of Ministry of Education, Research, Development and Youth of the Slovak Republic (project 2/0039/26) and the Doktogrant of Slovak Academy of Sciences (project no APP0578) is also gratefully acknowledged.